\documentclass[aip,reprint]{revtex4-1}
\pdfoutput=1

\usepackage{graphicx}
\usepackage{amsmath}
\usepackage{amssymb} \usepackage{eurosym}

\begin{document}


%\preprint{}

\title{Migraine generator network and spreading depression dynamics as
neuromodulation targets in episodic migraine}

\author{Markus A. Dahlem} \email[]{dahlem@physik.hu-berlin.de}
\homepage[]{https://sites.google.com/site/markusadahlem/}
%\thanks{}

\affiliation{Institute of Physics, Humboldt-Universit\"at zu Berlin, Berlin, Germany}


\date{\today}

%%%%%%%%%%%%%%%%%%%%%%%%%%%%%%%%%%%%%%%%%%%%%% % %% % The Abstract begins here
%% % %% % The Section headings here are those for  % % a Research article
%submitted to a        % % BMC-Series journal.                      %%  % %% %
%If your article is not of this type,     %% % then refer to the Instructions
%for       %% % authors on http://www.biomedcentral.com  %% % and change the
%section headings          %% % accordingly.  %%   %
%%% %%%%%%%%%%%%%%%%%%%%%%%%%%%%%%%%%%%%%%%%%%%%%

\begin{abstract} Migraine is a common disabling headache disorder characterized
by recurrent episodes sometimes preceded or accompanied by focal neurological
symptoms called aura. The relation between two subtypes, migraine without aura
(MWoA) and migraine with aura (MWA), is explored with the aim to identify
targets for neuromodulation techniques. To this end, a dynamically regulated
control system is schematically reduced to a network of the trigeminal nerve,
which innervates the cranial circulation, an associated descending modulatory
network of brainstem nuclei, and parasympathetic vasomotor efferents.  This
extends the idea of a  migraine generator region in the brainstem to a larger
network and is still simple and explicit enough to open up possibilities for
mathematical modeling in the future. In this study,  it is suggested that the
migraine generator network (MGN) is driven and may therefore respond
differently to different spatio-temporal noxious input in the migraine subtypes
MWA and MWoA. The noxious input is caused by a cortical perturbation of
homeostasis, known as spreading depression (SD). The MGN  might even trigger SD
in the first place by a failure in vasomotor control.  As a consequence,
migraine is considered as an inherently dynamical disease to which a linear
course from upstream to downstream events would not do justice.  Minimally
invasive and noninvasive neuromodulation techniques are briefly reviewed and
their rational is discussed in the context of the proposed mechanism.
\end{abstract}

\maketitle

%%%%%%%%%%%%%%%%%%%%%%%%%%%%%%%%
%\section*{Authors contributions}
%    Text for this section \ldots

{\bf Migraine is characterized by recurrent attacks of  moderate to severe
headaches sometimes preceded by visual, sensory, motor, or language
disturbances. These so-called migraine aura symptoms are caused by a chemical
imbalance in the brain that lasts usually not longer than one hour, while the
longer lasting headaches originate from neural activity in the brainstem.
Both have been observed with non-invasive imaging. In this study, it is proposed
how these events are linked together and that migraine pathophysiology should
be considered as a dynamical network phenomenon.}

% inflammatory mediators

\section{Introduction}

Migraine is one of the most prevalent neurological disorders with a life-time
prevalence of about 14\%.  Migraines can cause substantial levels of
disability, ranking on a disability scale from 0.0-1.0 at 0.7\cite{STO07}.  A
migraine attack can have up to four distinct stages: a prodromal phase with
difficulty concentrating, yawning, and fatigue;  an aura phase with all kinds of
sensory disturbances; the headache phase, often unilateral and throbbing; and
finally a  postdrome of tiredness, difficulty concentrating, persistence of
sensitivity to light and noise.  Yet, despite a distinctive clinical picture,
migraine continues to be underdiagnosed and undertreated. It is estimated that
the cost for the US and European economies sum up to US\$19.6 billion and
\euro{}27 billion a year, respectively.

There are several subforms of migraine, yet the two main subtypes are
migraine without aura (MWoA) and migraine with aura (MWA). MWA comprises, for
instance, six subforms, one even without headache.  While for most migraine
sufferers, the head pain is the key manifestation of their disorder, the
migraine aura is also a distinctive feature: often visual hallucinations,
language problems, motor weakness and other short-duration (5-60min) changes to
sensory modalities or cognitive functions. These symptoms occur only in MWA,
i.e., in about 30\% of the cases and usually precede or sometimes accompany the
headache\cite{HAN12a}. Why not in all attacks? This raises a related more
general question: Is the pain mechanism in MWoA and MWA the same?  If so, what
drives the activation of neuronal pain pathways and how is this related to the
cause of aura symptoms?  And what determines which sensory modality or
cognitive function is affected?  These interlinked questions about migraine
pathophysiology remain open or are only unsatisfactorily answered in current
migraine theories. 

The migraine generator and the spreading depression (SD) theory of migraine
both focus on distinct pathophysiologic events, yet they are not mutually
exclusive theories of the cause of episodic migraine. The issues raised above
might be resolved, if both theories can be unified.

From a mathematical point of view, migraine pathophysiology involves sudden
dynamical transitions, because although migraine is a chronic neurological
disorder it is characterized by recurrent episodes.  In these transitions both
temporal rhythms and spatio-temporal patterns change, which allows only one
conclusion: migraine is a dynamical disease; a term  coined to identify
diseases that occur due to an abrupt change, usually a bifurcation.  In fact
originally, dynamical diseases are defined as a situation where sudden changes
bring the system through a bifurcation\cite{MAC87}.  But in a wider sense, if
we have---or aim for---an understanding of the pathophysiological mechanism in
terms of equations of motion and low dimensional phase space structures, we may
call it a dynamical disease.  The sudden change may also originate from other
dynamical phenomena than bifurcations, such as intermittency, noise driven
excitation behavior, etc.. These various points that lead to sudden
transitions may also be summarized as ``tipping points''.

As a consequence, migraine pathophysiology needs to be considered as a problem
at the interface of clinical neurology and  applied nonlinear science.  To this
end, the migraine generator theory must be extended from a theory of brainstem
regions as the origin of attacks to a network concept where the origin is not
localized but a dynamical transition into dysfunctional
control\cite{WEI95a,WEL01,FOX05,BOR12}. In Sec.~\ref{sec:mg}, an extended
but still reduced migraine generator network (MGN) will be introduced. This
network is connected to the cortex, where the spatio-temporal patterns of SD
occur.  SD itself can be describes by a macroscopic continuous limit of
cortical tissue in terms of reaction-diffusion systems (Sec.~\ref{sec:sd}). How
do these two dynamical systems interact in migraine, that is, how is the
central pattern generator network coupled to the cortical reaction-diffusion
system?

A possible relation is proposed in this theoretical study. The pivotal role
lies in the driving input of the MGN. The MGN is a system that compromises  two
physiological control subsystems, the trigeminovascular system for vasomotor
control and associated descending modulatory brainstem system for pain control.
The vasomotor system may even lead to the ignition of cortical SD in the first
place, which then in turn drives the MGN. We do not consider the cause of the
ignition of SD but suggest SD---once ignited---drives the MGN and this systems
responds differently to different spatio-temporal signatures of SD, that is,
characteristic noxious action of SD leads to the subforms MWA and MWoA.  In
this study, only the driving mechanism is considered in some detail in
Sec.~\ref{sec:sd}.

A mathematical model of this dynamical response is beyond the scope here,
though the MGN itself is presented as a reduced scheme explicit enough to
support the conclusions for neuromodulation techniques
(Sec.~\ref{sec:neuromodulation}).  In general, this dynamical response is a
form of central sensitization, which refers to a transition in a pain
network---part of the {\em central} nervous system---after which it responds
with increased sensitivity to noxious stimuli (hyperalgesia) and even
non-noxious stimuli (allodynia). In episodic migraine, the central
sensitization is probably  one of second-order neurons\cite{BUR05a}.  Chronic
migraine, in contrast, central sensitization is probably a dynamical transition
in the pain matrix (see below), which also leads to an enhanced response to
peripheral input but is not considered here.  

In Sec.~\ref{sec:neuromodulation}, minimally invasive and noninvasive
neuromodulation techniques are briefly reviewed. These techniques target neural
structures in the peripheral and central nervous system that have been
introduced in Sec.~\ref{sec:mg} and Sec.~\ref{sec:sd}.  The aim of these
techniques is to abort migraine. I discuss their rational in the context of the
proposed driving mechanism of the MGN by SD dynamics. 

In conclusion, noninvasive techniques for episodic migraine should be most
effective in transcranial stimulation with three distinct modes of action, for
(i) the prodromal phase, (ii) the ignition and (iii) the acute phase of
migraine aura. Furthermore, one mode of action in transcutaneous stimulation
(iv) in the acute pain phase of the episodic attacks. Further quantitative
approaches are needed to find optimized stimulation protocols for these
situations based on the proposed mechanism.

%:sec\pagebreak
\section{Migraine generator network} 
\label{sec:mg}

In several respects, pain is different from other sensations. For instance,
normal sensory pathways take a chain of three neurons to get from receptors, e.g.,
sensitive mechano\-receptors, to the primary sensory cortical area.  In pain
traffic, there are more intermediate nuclei and a cascade of descending control
mechanisms is essential for defining the pain experience\cite{HEi09a}.
Therefore a simple order of neurons on the way to the cortex is questionable.
Furthermore, instead of a single, i.\,e., spatially confined, primary cortical
area as the first target in the cortex, a neural correlate of pain perception
seems to be found in spatially segregated activity patterns of a widespread
network of cortical areas.

\begin{figure}[t] \begin{center}
\includegraphics[width=\columnwidth]{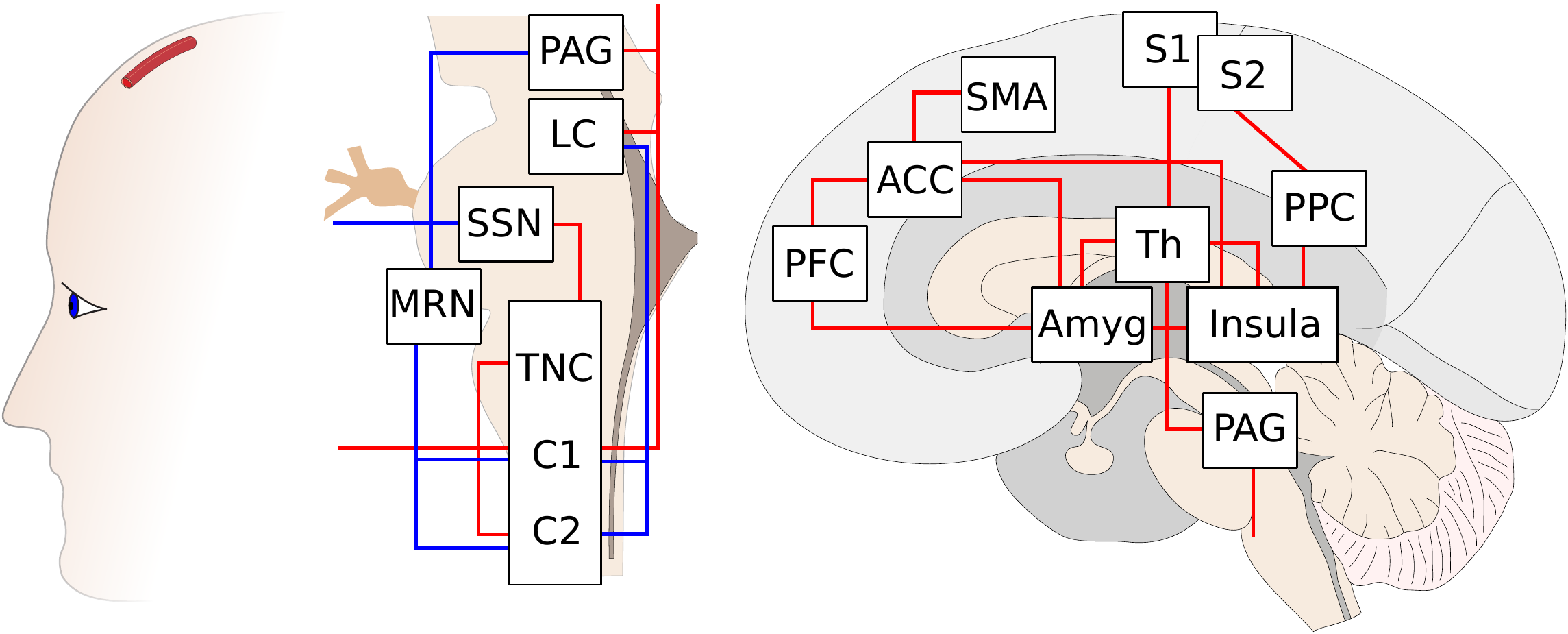}
\caption{\label{fig:painFormation}(color online) Networks related to pain. (a)
Cranial circulation (CC): connected by first-order perivascular pain-sensitive
neurons via the trigenimal ganglion (TG) to the trigeminocervical complex (TCC,
see b) and receives also input from the superior salivatory nucleus (SSN). (b)
Brainstem: trigeminal ganglion (TG); trigeminal nucleus caudalis (TNC) and
C1-C2 regions compromise the trigeminocervical complex (TCC); rostral
ventromedial medulla (RVM), locus coeruleus (LC), ventrolateral periaqueductal
grey (vlPAG). (c) Pain matrix\cite{MAY06a}: thalamus (TH); anterior cingulate
cortex (ACC); amygdala (Amyg); (PFC); primary (S1) and secondary (S2)
somatosensory cortices; supplementary motor area (SMA); prefrontal cortex
(PPC); insula cortex (Insula).} \end{center} \end{figure}

Whether there is a network specific for pain perception, the ``pain
matrix''---a term that usually also includes diencephalic structures in the
forebrain such as the thalamus, see Fig.~\ref{fig:painFormation}---or not and
instead this network actually is a largely unspecific sensory ``neuromatrix'',
is still debated\cite{IAN10}.  In any case, this matrix seems not to be an
optimal target  in episodic migraine treatment, e.g., for neuromodulation
techniques, because its activity patterns are probably truly downstream events
(but cf.~chronification of pain, i.\,e., $>$15 headache days per month over a 3
month period, at least 8 migrainous, and absence of medication overuse).   

The question of the most upstream event is more difficult (see Discussion), and
it may only make sense to talk about upstream events in the form of transitions in
dynamical states of the migraine generator and spreading depression
(Sec.~\ref{sec:sd}).

\begin{figure}[b] \begin{center}
\includegraphics[width=0.5\columnwidth]{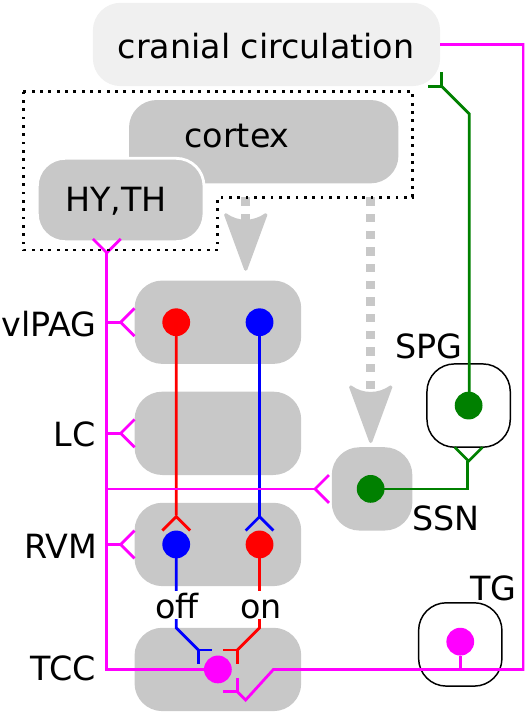}
\caption{\label{fig:painNetwork} (color online) Schematic diagram of pathways in the
migraine generator network (MGN)\cite{PIE03,AKE11,MEN11a}.  } \end{center} \end{figure}

While it is worth illustrating all nuclei of this network in their
respective anatomical place (in approximation), it is still necessary to
identify a schematically reduced network of these nuclei, their neural
subpopulations, and a vasomotor loop that is explicit enough to convey the
regulatory control in this network structure amenable for mathematical
analysis, see Fig.~\ref{fig:painNetwork}. For this reason, the following brain
regions and their respective functions are briefly introduced.

\subsection{The trigeminovascular system}

The trigeminovascular system, which is key to understanding the network structure
of the MGN, is a largely bidirectional system. Notwithstanding, it is
natural to consider as a starting point, if only for convenience, the
nociceptor and the afferent projection of the first-order neuron, that is, the
innervation of both dura mater and large intracranial vessels by the ophthalmic
division of the trigeminal nerve.  This innervation of the cranial circulation
(CC) terminates in the central trigeminal nucleus caudalis (TNC) and the C1-C2
regions of the cervical spinal cord, together known as the trigeminocervical
complex (TCC), see Fig.~\ref{fig:painFormation}.  

Beyond the TCC, nociceptive traffic ascends by activated second-order neurons
to further brainstem nuclei: the rostral ventromedial medulla (RVM) including
the nucleus raphe magnus (NRM) in the medulla; the locus coeruleus (LC) in the
dorsolateral pons (both hindbrain); and the ventrolateral periaqueductal grey (vlPAG) in
the tegmentum (midbrain). Furthermore, the TCC also ascends  nociceptive
information directly to the hypothalamus (HY) and thalamus (TH) with
second-order neurons along the trigemino\-hypothalamic tract and the
trigemino\-thalamic tract (also know as quintothalamic tract), respectively. TH
and HY are structures in the diencephalon of the forebrain from where
third-order neurons ascend to the cortex (CTX), the final telencephalic station
within the forebrain. 

For the sake of completeness, one other pathway of the trigeminovascular
systems should be mentioned.  There is a peripheral vasomotor connection from
efferents of the superior cervical spinal cord to the cranial
circulation\cite{PUR78}.

\subsection{Cortical and diencephalic influence}

Before we come to the descending modulatory network of the brainstem, let us
briefly mention descending pathways from the pain matrix. Cortical projections
to the periaqueductal grey (PAG) arise from the medial network of the prefrontal cortex (PFC) or
cortical areas that are closely related to this network\cite{PRI06}. The
interest here is obviously the part which belongs to the MGN, in
particular the ventrolateral (vl) PAG. The vlPAG evokes passive coping
reactions (withdrawal, quiescence)\cite{KEA01} consistent with behavior
exhibited by migraine patients. Since such coping behavior is also readily
elicited in animals with the forebrain being disconnected from the brain
stem\cite{KEA01}, a less important role of these descending pathways for the
dynamics of the MGN is expected.  

Therefore, we do not include the detailed areas or even a closed feedback loop
through the pain matrix in the attempt to construct a basic structure of the
MGN in Fig.~\ref{fig:painNetwork}, though this input is indicated by a dotted
grey arrow.  To support this further, it could be shown with non-invasive
imaging that brainstem activation in humans persists in migraine even after
pain relief and reduction of symptoms by acute treatment\cite{WEI95a}. However,
it might be that not PAG but nearby nuclei in the dorsolateral pons are causing
the activation\cite{BOR12}.    

Notwithstanding, the thalamus is certainly key for integration of
nociceptive inputs in migraine and the pain matrix for pain
sensation\cite{AKE11}. In particular to understand chronification of pain,
closed feedback loops  through these areas are probably important and these
diencephalic or cortical areas may even primarily be the ``upstream'' target
areas for a therapeutic rational in chronic migraine.

\subsection{Descending modulatory network}

Given that PAG is an area to which various cortical and diencephalic
structures project, it is important to note that PAG does not directly project
itself to the TCC. The descending projection from PAG modulate ON and OFF
neurons in RVM.  As indicated by the name, these ON and OFF neurons in the
PAG-RVM descending control system can be facilitatory and inhibitory depending
on how the balance shifts in either direction\cite{MEN11a}. A descending
facilitation of TCC can provide a mechanism for central sensitisation.  The
development of cutaneous allodynia suggests that the balance shifts in favor of
facilitation explaining the increased sensitivity of the skin to non-noxious 
stimuli in migraine\cite{BUR00}.

It has been questioned whether the PAG or rather the nuclei in the dorsolateral
pons, like LC, are responsible in this descending modulatory network and it was
suggested that ``[t]o advance the brainstem migraine generator theory from the
opinion phase to being evidence based, answers should be provided to questions
as: (a) How would increased activation in the PAG drive or produce
migraine?''\cite{BOR12}. Quantitative approaches in a form of a  mathematical
model similar to the gate theory of first-order neurons\cite{BRI89} may
contribute to this question. The important notion stressed here is that the
extended migraine generator network theory is essentially a dynamic network
theory.  The MGN is a central pattern generator, that is, a  neural network
that produces rhythmic patterned output perceived as pain without sensory
feedback. However, such feedback is still needed, as proposed here,  to
drive the transition into central sensitization. The explanatory power of a
mathematical approach lies in the predictive power, but it certainly also needs
the clinical testing to be evidence based.

\subsection{Vasomotor control}

The key vasomotor control is probably played by a parasympathetic cluster of
neuronal cell bodies located in a fossa (ditch) in the skull, the
sphenopalatine ganglion (SPG).  The simplest putative structure of the MGN
(Fig.~\ref{fig:painNetwork}) is completed by the SPG and its second-order
preganglionic neurons in the superior salivatory nucleus (SSN).  It was
suggested that migraine triggers typical for the prodromal phase either
activate or even originate in a number of brain areas whose projections
converge on the SSN (dotted grey arrow)  and which are functionally positioned
to produce migraine symptoms \cite{BUR05a}.  \\

%:sec\pagebreak
\section{Spreading depression theory of migraine} \label{sec:sd}

SD is essentially a slow (about 3mm/min\,=\,50$\mu$m/sec) reaction-diffusion
wave in gray matter tissue.  In the cortex, SD is accompanied by a pronounced
hemodynamic response of increased regional cerebral blood flow (hypermia) for
about 2min and a long lasting, $\sim$2h, decrease (oligemia)\cite{DRE11}.

The cortical tissue SD traverses is massively  perturbed in its ion
homeostasis\cite{SOM01}. If the ion flow across the membrane in this region
is taken to estimate relative changes in Gibbs free energy of the tissue,
SD reveals itself as  ``a twilight state close to death''\cite{DRE12}, i.\,e.,
during the peak of SD the cortical state is similar to the one in stroke or
ischemia. The ion flow causing this state is far more dramatic than in any
other neurological disorders, for instance, compared to the ion flow during
ictal epileptic activity. 

However, the cortex is not pain sensitive. It was therefore suggested that SD
may trigger the pain phase indirectly\cite{MOS93,BOL02}.

\subsection{SD and pain pathways}

In several animal models, possible pathways were investigated to see how SD might
cause pain by activating nociceptors in sensitive intracranial tissues and
subsequently activate the TCC\cite{EIK08}. In short, it was suggested that
during SD the blood--brain barrier becomes more permeable, allowing glutamate,
potassium ions, hydrogen ions, nitric oxide and other noxious substances or
inflammatory mediators to diffuse from the surface of the cortex into the
meninges where they activate nociceptors of the cranial circulation. Although
this mechanism has been criticized\cite{TFE11}, the animal models of migraine
show that SD activates the first-order neurons in the trigeminal nucleus
caudalis\cite{MOS93,BOL02,ING97,MOS98}.  In a newly published article, it was
found that SD can trigger headache by activating neuronal Panx1 channels, a
megachannel that is a structural component of gap junctions\cite{KAR13}.  It is
furthermore discussed, whether nuclei in the MGN in turn can cause
spreading depression in the cortex.  For a summary, see Ref.~\cite{AYA10}.

\subsection{SD, aura, and silent courses}

Even the apparently solved question of how SD relates to  the aura phase has still
some merit.  It is clear that the aura is caused by SD.  At least after
successfully imaging the blood flow in migraine aura, it became indisputable
that SD is the electrophysiological correlate of migraine
aura\cite{OLE81,LAU87,HAD01}. In fact, already in 1945, Le\~ao wrote: ``The
latter disease [migraine] with the marked dilatation of major blood vessels and
the slow march of scotomata [sensory blindness] in the visual or somatic
sensory sphere is suggestively similar to the experimental phenomenon here
described [SD], in spite of the fact that known scotomata are still felt to be
vasoconstrictor in nature''\cite{LEA45}. It was not known in 1945 that in
migraine, a vasoconstrictor response (oligemia) is also present.

%Nonetheless, it is remarkable that it took that long to establish this
%relationship. 

To date, what remains disputed is whether SD is even present but clinically
silent in the migraine subtype MWoA. This is referred to by the contested
notion of ``silent aura''\cite{AYA10}. If true, it is SD which stays
silent, the aura is simply not present. Therefore ``silent SD'' is a term less
perplexing.  Anyway, such silent SD courses are supported by (a) a well-documented
case of blood-flow changes that were likely the result of SD observed in a
spontaneous migraine headache without aura\cite{WOO94} and (b) the fact that SD can cause
pain and MWA and MWoA are likely to share the same pain mechanism.\\

\begin{figure}[t] \begin{center}
\includegraphics[width=0.7\columnwidth]{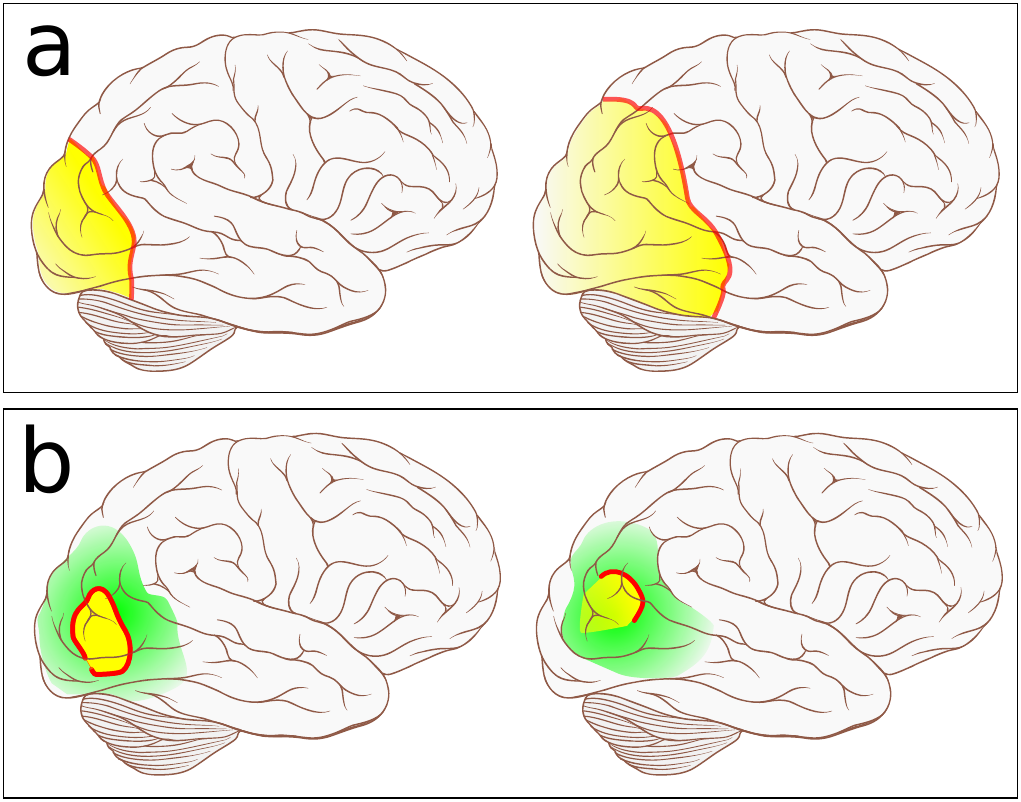}
\caption{\label{fig:csdEngulfingVsLocalized} (color online) Paradigms of SD
pattern in migraine. Narrow band of excitation (red), refractory zone (yellow),
disperse neurovascular feedback (green). (a) Typical textbook illustration: SD
engulfing large parts of one hemisphere; adopted from Ref.\cite{LAU87}. (b)
Localized SD wave in a full-scale attack\cite{HAD01,DAH08d}.} \end{center}
\end{figure}

To resolve these questions about the relation of SD to the pain and aura phase,
we recently suggested that which of the two migraine subtypes develops, i.\,e.,
MWoA or MWA, is dependent upon the spatio-temporal pattern (shape, size, and
duration) of SD\cite{DAH12b}.  This does not rule out the possibility that SD
itself is triggered by events in the brainstem and therefore we contribute only
little to the questions of the most upstream event. However,  the crucial part
missed in the current controversy might be the spatio-temporal pattern of
the cortical tissue being recruited into a depleted state. All too often, it is
completely ignored that SD usually propagates in the human gyrified cortex as a
discontinuous wave segment as observed with fMRI in migraine and reported by
patients as visual field defects\cite{HAD01,DAH04b,DAH08d}, see
Fig.~\ref{fig:csdEngulfingVsLocalized}b.  In animal models, SD is engulfing the
whole lissencephalic cerebral hemisphere but already in gyrencephalic cat brain
only the first SD waves do this and succeeding secondary SD waves, which
propagate in relative refractory, i.\,e., less susceptible, cortex, often
remained within a confined origin\cite{JAM99}.

\subsection{Localized SD waves}

Both fMRI data and reports on visual field defects\cite{HAD01,DAH04b,DAH08d} clearly show
that SD is not only spatially confined in humans, but SD is a discontinuous
wave---a particle-like wave. A  particle-like wave is a wave front with two
open ends, sometimes also called dissipative soliton\cite{AKH08}. The existence
of such localized waves is a change in the pattern forming paradigm indicating
a fast diffusing second inhibitor related to the Turing bifurcation.

From a clinical perspective, migraine aura symptoms are focal symptoms.  If SD
were to spread out in all directions, the affected cortical area would increase quickly and
symptoms should become more generalized.  Localization does not merely restrict the
symptoms caused by SD; but localization also allows SD to assume a much larger
variety of spatio-temporal patterns compared to a continuous circular wave
front. This is reflected in the variety of the courses of focal aura
symptoms\cite{VIN07} and maybe also in the fact that SD could stay silent under
certain conditions on the localised pattern (Sec.~III~B,F) and, furthermore,
that SD  does not cause pain under other conditions (Sec.~III~F-G).  

For these reasons, it is an essential feature that SD stays localised even when
it propagates a long distance in one direction.  In contrast, if SD spread out
in all directions, as to date shown in all modern migraine textbooks (published
or revised in the last 5 years), the shape would essentially be circular
(cortical folding has some influence) and, furthermore, size and duration both
would depend on a single parameter, the radial distance SD travels. The course
that is usually sketched starts from the occipital pole of the cortex and SD
propagates in a full-scale MWA attack up until the frontal lobe, see
Fig.~\ref{fig:csdEngulfingVsLocalized}a. If SD invaded the complete occipital,
temporal, and parietal lobe, this would not match with the reported courses of
neurological symptoms during the aura phase, not even in a full-scale attack,
cf.~Ref.~\cite{VIN07}, and it is also implausible that only a subset of the
nearly completely depleted cortical tissue results in some form of clinical
manifestation.

A simple calculation illustrates the different scaling behavior of events.  In
a full-scale MWA attack, the aura phase can last up to one hour. This
corresponds to a length $l$ of $l_1=18$cm cortex being paced. Even if we halve
this distance $l_2=9$cm--- either because SD is slower or lasting only 30min
and symptoms reverberate for a full 60min---this would still correspond to 20\%
(80\% for $l_1=18$cm) of the surface area of one cerebral hemisphere. In
contrast, only 1.5\% (3\% for $l_1=18$cm) of the surface are affected in a
full-scale MWA, if the particle-like SD wave is on average 2cm wide. Note that
the progressive development of aura symptoms is, except for the ignition phase
of SD, probably  mostly due to the fact that SD starts in regions of high
cortical sensory magnification and travels into regions with lower cortical
magnification\cite{DAH12a}.

\subsection{Generic reaction-diffusion model}

In accordance with the noninvasively imaged SD progression and reported visual
field defects, we proposed a model in which in each attack a particular pattern
is formed by a discontinuous wave segment that spreads out only in one
direction determined by the initial conditions\cite{DAH12b} and guided by
the folding pattern (unpublished results).  The ignition patterns that trigger these simulated
attacks are given by an assumed hyperactivity  in a cortical feature map of the
visual cortex used as initial conditions for the model.   These feature map
patterns were also suggested to be hyperactive during SD causing visual
hallucinatory percepts\cite{DAH00a}.

In this generic reaction-diffusion mechanism of excitable media, the resulting
pattern forming processes following these initial conditions are described
in abstract terms of activator-inhibitor kinetics of two variables $u$ and $v$,
respectively. In fact, simple activator kinetics of $u$ date  back to a
mechanism of SD described in 1963 by Grafstein, Hodgkin and Huxley
(GHH)\cite{GRA63}.  The GHH model assumed $u\equiv[K^+]_e$, i.\,e., the
extracellular potassium ion concentration to be the activator $u$. We extended
this scheme (for details of the full model see Ref.~\cite{DAH12b}):  

\begin{eqnarray} \frac{\partial u}{\partial t} &=&  u -\frac{1}{3}
u^3 -v + D\nabla^2u,   \label{eq:u}      \\ \frac{\partial v}{\partial t} &=& \varepsilon \left( u
+ \beta + K \int H\left(u\right)\,\textrm{d} x\textrm{d} y \label{eq:v} \right),
\end{eqnarray} with $D$ the diffusion coefficient, two parameters $\varepsilon$ (time scale separation) and $\beta$ (threshold), 
$H$ being the Heaviside step function (discontinuous
function whose value for negative (positive) arguments is 0 (1) and 0.5 for the
argument zero). The integral term is therefore a surface area measure of the medium
being in a state with $u$ larger than zero that can globally increase the threshold. 

Both activator $u$ and inhibitor $v$ are lump variables.
Unlike in the original GHH model, it is not necessary or even possible to
identify particular physiological quantities, like $[K^+]_e$, with them.
Instead, the activator $u$ should be viewed as the bistable energy state. One
stable fixed point corresponds to the maintenance of homeostasis far from
thermodynamic equilibrium and the other stable fixed point to the state where
the cellular Nernst reversal potential are depleted (thermodynamic
equilibrium). In the presence of an inhibitor $v$, which was not included in
the GHH model, the second stable fixed point becomes a transient state. In
other words, the inhibitor $v$ is a recovery variable. It is related to ion
pumps that drive the tissue far from thermodynamic equilibrium and (re)charge
Nernst reversal potentials.  We set the parameters as follows: time scale
separation $\varepsilon=0.04$, the threshold $\beta=1.32$, and the mean field
coupling $K=0.003$. 

\begin{figure}[b] \begin{center}
\includegraphics[width=\columnwidth]{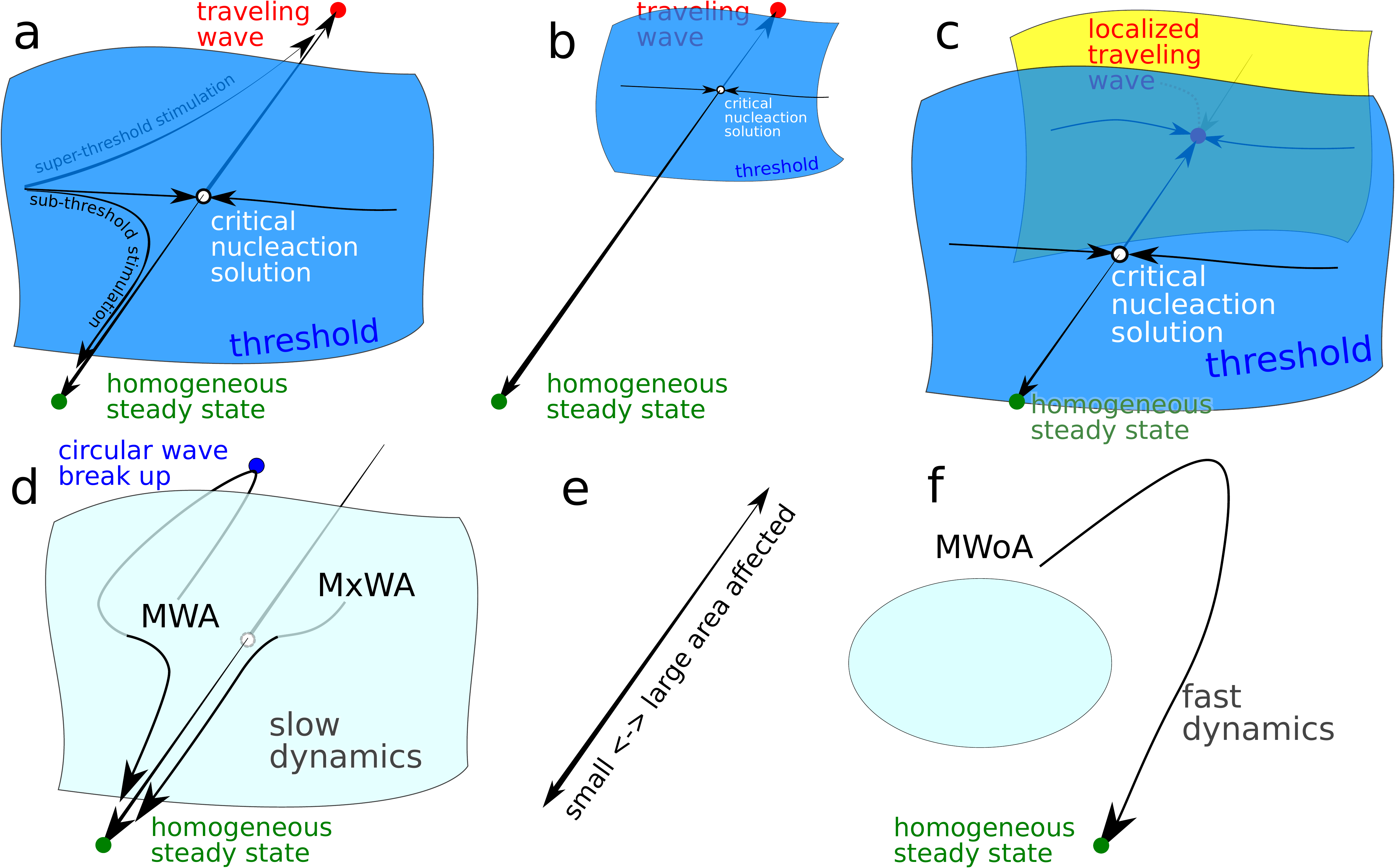}
\caption{\label{fig:excitableMediumPhasePortraitMeanField-all}(color online) Schematic orbit
structures in phase space\cite{DAH12b}, see text. (a) Caricature of excitable
media. (b) Basin of attraction change when local excitabilty parameter change.
(c) Fast secondary inhibition creates localized structures.  (d) Ghost
behaviour when traveling wave solution (node) collides with nucleation
(saddle): the two trajectories are associated with  migraine with aura (MWA) 
and with migraine with typical aura without headache  (MxWA). (e) Direction of cortical area
affected by SD. (f) Trajectory of a circular wave associated with migraine
without aura (MWoA). } \end{center} \end{figure}

The details of the model are discussed elsewhere\cite{DAH04b,DAH12b}. In
short, for $K=0$, the bistable activator kinetics with diffusion and the linear
inhibitor kinetics (immobilised) provide the simplest reaction-diffusion model
of activator-inhibitor type for the cortex as an excitable media exhibiting
patterns as seen during SD in animal models. The homogeneous steady state is a
stable solution representing the healthy cortical state.  A critical neuronal
mass is needed to start ignition and support sustained propagation of a Gibbs
free energy--depleted state, which is associated with the state of
SD\cite{DRE12}.  For $K=0$, this depleted SD state would engulf all the tissue
after an initial ignition.  The suggested pattern forming mechanism of
localized SD is closely related to the critical mass needed for ignition,
called a critical nucleation solution, see
Fig.~\ref{fig:excitableMediumPhasePortraitMeanField-all}a.

This critical nucleation is controlled for $K\neq0$ by a widely spread out
inhibitory feedback,
Fig.~\ref{fig:excitableMediumPhasePortraitMeanField-all}b,c. This inhibitory
feedback is approximated by a mean field term $K \int
H\left(u\right)\,\textrm{d} x\textrm{d} y$, a term proportional to the depleted
surface area region in the cortex.  The physiological significance of this last
term in the rate function of the lump variable $v$ is associated with the
neurovascular feedback (e.g., as observed during hypermia of increased blood
flow\cite{OLE81,HAD01}) in SD. For the sake of simplicity, we chose it to be
proportional to the surface area of the depleted state, although the
neurovascular feedback is mainly driven by the neuronal hyperactivity in the
rise of the SD front. Therefore, a measure that is  proportional to the length of
the SD front may be more appropriate. However, we tested $K \int
H\left(u\right)H\left(1-v\right)\,\textrm{d} x\textrm{d} y$, which is an
approximated measure of the length of the SD front---the activator must be
in the excited state while the inhibitor is still low, as characteristic for
the hyperactivity in the rising front---with similar results.

\subsection{Transient waves and slow dynamics}

\begin{figure}[t] \begin{center}
\includegraphics[width=\columnwidth]{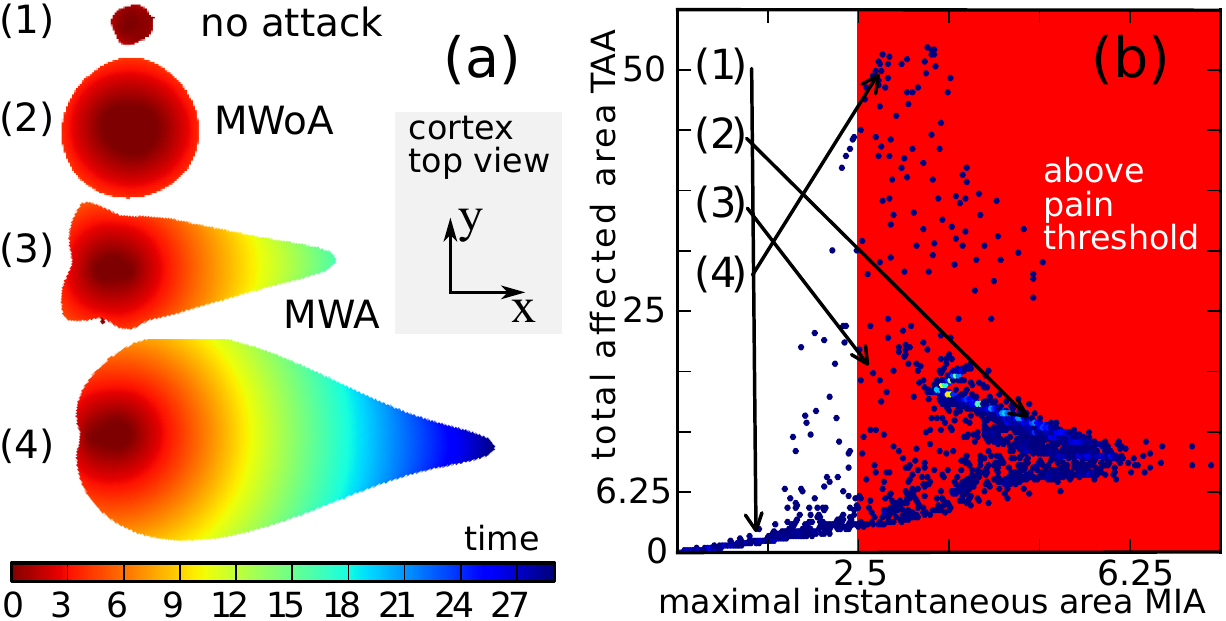}
\caption{\label{fig:bestiaMIAvsTAA}(color online) (a) Spatio-temporal signatures of SD as
obtained from a reaction-diffusion model with mean field coupling,
Eqs.~(1)-(2); cf. Fig.~4E in Ref.\cite{HAD01}.  The model parameters are chosen
such that only transient waves exist\cite{DAH12b}, note that we have scaled the
dimensionless model such that the units of time are roughly corresponding to
minutes, and surface area to cm$^2$ and the wave spreads with about 3mm/min.
Four stereotypical courses are depicted following a local perturbation of a
homogeneous steady state: (1) subthreshold, SD dies out quickly without initial
spread; (2) superthreshold perturbation, SD dies out after a few minutes, if
its front does not break open, corresponding to migraine without aura (MWoA);
(3)-(4) superthreshold perturbation, SD propagates for up to 30min ($\sim$15 in
(3), $\sim$25min in (4)), if the SD front breaks open, corresponding to
migraine with aura (MWA).  (b) Statistical analysis of 8000 events plotted over
MIA and TAA, see text.  } \end{center} \end{figure}

Changing a parameter of this model without the mean field inhibitory feedback
($K=0$), essentially varies the basins of attraction of the existing solutions,
e.\,g., making the medium less susceptible by increasing the basin of
attraction of the homogeneous steady state,
Fig.~\ref{fig:excitableMediumPhasePortraitMeanField-all}b. For $K\neq0$, the
traveling wave solution itself is changed such that the traveling wave solution
becomes localized, as described in the previous section,
Fig.~\ref{fig:excitableMediumPhasePortraitMeanField-all}c.  For sufficiently
large $K$, this stable localized solution collides with its nucleation solution
(saddle), which leads to well known phenomenon of slow dynamics named ghost
behavior after a saddle-node bifurcation,
Fig.~\ref{fig:excitableMediumPhasePortraitMeanField-all}d. This is the origin
of the transient nature of the SD signatures in Fig.~\ref{fig:bestiaMIAvsTAA}
that we propose to lead to migraine without aura, migraine with aura, and
migraine with typical aura and without headache (MWoA, MWA, and MxWA,
respectively). 

In particular, MWA, and MxWA correspond to trajectories that pass the ghost of
the slow manifold (lightblue in
Fig.~\ref{fig:excitableMediumPhasePortraitMeanField-all}d). In contrast,
solutions that are shorter lasting but with larger MIA
(cf.~Fig.~\ref{fig:excitableMediumPhasePortraitMeanField-all}e) do not pass the
ghost of the slow manifold
Fig.~\ref{fig:excitableMediumPhasePortraitMeanField-all}f; they correspond to
MWoA.

In Sec.~IV\,D, this is further
discussed in the context of the therapeutic strategy for neuromodulation. 

 The macroscopic features of SD are well described by this model. The
affected cortical area will not grow in time and SD stays localized while
propagating (for further discussion of the usefulness of such a toy model, see
Sec.~\ref{sec:ModelBasedControl}). It is worth noting that it is not without
irony that both terms in the name of SD are rather misleading: the important
neural activity is a hyperactivity that only in the depleted phase becomes
``depressed'', and this cortical state is only in the ignition phase
``spreading'', the sustained main process is a traveling wave.

\subsection{Link to migraine subforms}

We performed statistical analysis of the spatio-temporal development governed
by the generic reaction-diffusion model in Eq.~(1)-(2) that followed 8000 different 
natural initial perturbations of local hyperactivity as SD triggers. These
perturbations were introduced by an increase of the resting value of $u$. 
Based on that, we predict that certain features related to shape, size, and
duration of SD (Fig.~\ref{fig:bestiaMIAvsTAA}a) determine the aura phase while
others determine the pain phase in migraine. 

It is important for further discussion that we firstly divide the
subtype MWA into its two subforms ``typical aura with migraine headache (1.2.1,
in the international classification)'' and ``typical aura without headache
(1.2.3)'', which we refer to as MxWA. Therefore, in the remainder of the
text, we refer to 1.2.1 as MWA, i.e., we explicitly assume the presence of
headache.  MxWA is estimated to account for about 5\% and MWA for 25\% of the
30\% of the subtype migraine with aura (1.2). The number of unreported
cases of MxWA might be quite large, though.  

Together with MWoA, three combinations exist.  We predict that each migraine
subform has a specific signature of SD.  Characteristic for this signature is
not only the spatio-temporal pattern of the course of SD in the gyrified human
cortex but also the hemodynamic response and the amount of noxious
substances that diffuse into the meninges. The cortical folding pattern
should also influence this signature. We are currently performing a detailed analysis
of this influence.

 \begin{figure}[t] \begin{center}
\includegraphics[width=0.75\columnwidth]{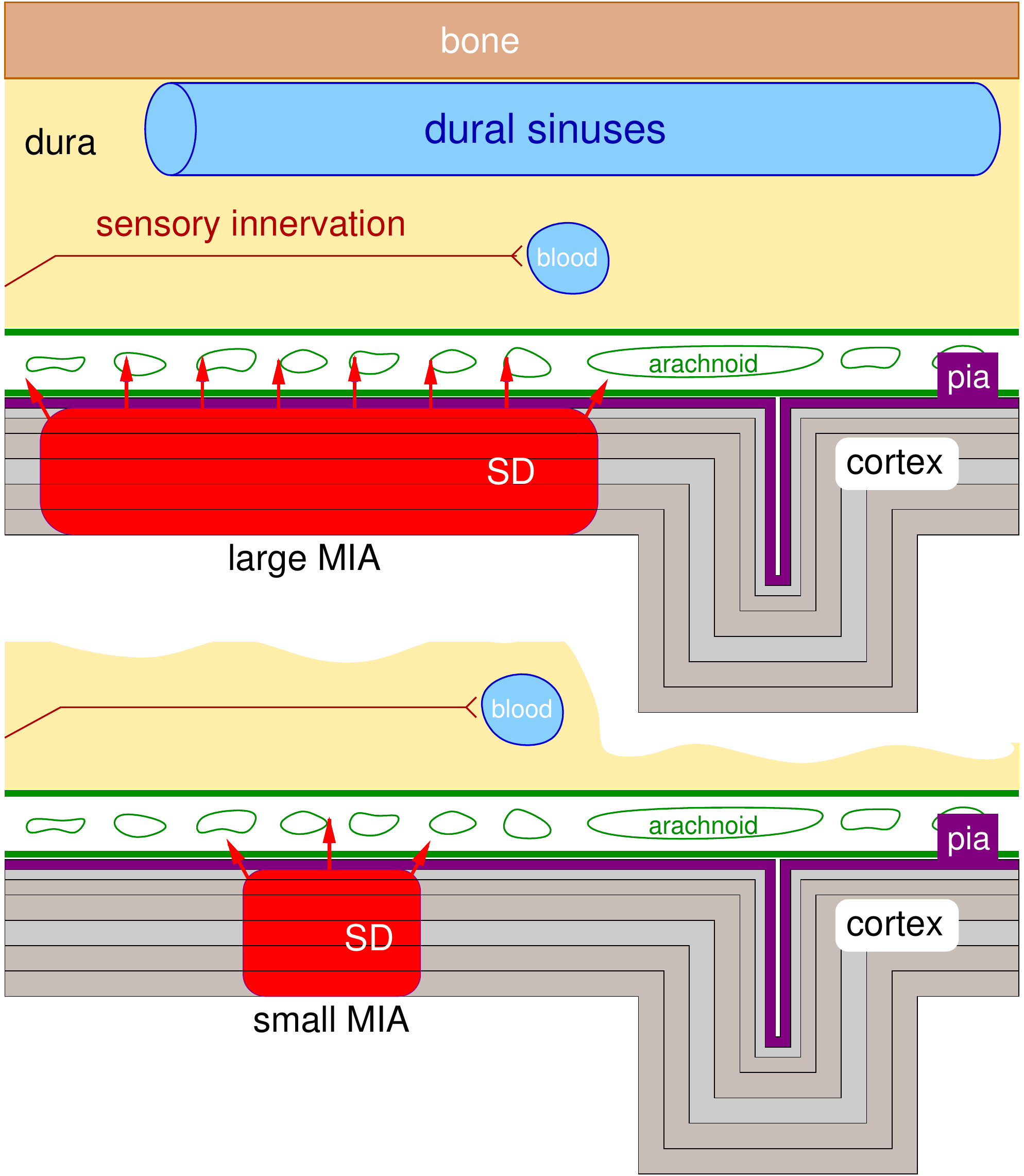}
\caption{\label{fig:cortexMeningesSDBestia}(color online) Schematic
representation of a cortical cross section with meninges and skull. Top: large
MIA; bottom: lower MIA, not to scale; cortical thickness is at the occipital
lobe about 2mm; diameter of MIA pain threshold is approximately 1.78cm.}
\end{center} \end{figure}

In particular, we predict that a sufficiently large surface area must be
instantaneously affected and thus depleted by SD to lead into the pain phase in
migraine.  Therefore, we propose that if  the maximal instantaneously affected
area (MIA, Fig.~\ref{fig:bestiaMIAvsTAA}b) is too small, the cascade of
subsequent events causing sustained activation of trigeminal afferents is not
initiated.  The rational behind this is that the transmission of substances
(noxious or inflammatory mediators) in the direction perpendicular to the
cortical surface into the pain sensitive meninges should be significantly
convergent to reach noxious threshold concentration and initiate central
sensitisation of second order neurons, Fig.~\ref{fig:cortexMeningesSDBestia}.
While the flow driven by small MIA is sufficiently diluted and therefore
tolerated. This makes a prediction that can be tested by noninvasive imaging.

Furthermore, only if SD assumes a spatio-temporal form that is lasting long
enough ($>$5min) and therefore also propagating farther ($>$1.5cm), SD will
cause noticeably aura symptoms. In fact, this is not a prediction. It merely
reflects the diagnostic criteria of migraine aura given by the International
Classification of Headache Disorders:  ``focal neurological symptoms that
usually develop gradually over 5-20 minutes''. So any neurological events that
last less than 5min are usually not diagnosed as migraine aura.   

The pain phase  is completely suppressed in cases of MxWA, if the ignition is
not causing a large MIA. Such a temporal evolution in Eq.~(1)-(2) is expected
for initial perturbations that are just large enough to be suprathreshold
stimulations and at the same time are asymmetric to lead directly into a
discontinuous wave (Fig.~\ref{fig:excitableMediumPhasePortraitMeanField-all}D)
that is slowed down. As a consequence from the model, MxWA   are particularly
long lasting auras, again a prediction that can be tested.

\subsection{Pain threshold determined by SD
reaction-diffusion model}  

The distribution of spatio-temporal SD patterns that results from local
perturbation led us to suggest a pain threshold determined by MIA. This value
can be estimated to be about MIA$\approx2.5$cm$^2$.  In other words, if SD is
never depleting the transmembrane gradients in cortical tissue over a 
surface area of more than 2.5cm$^2$ (at any given time), the pain cascade is not
initiated. This corresponds to a circular area with a diameter of about
$d\approx1.78$cm or 0.1\% of the total cortical surface. The course (2) shown
in Fig.~\ref{fig:bestiaMIAvsTAA} has a MIA of about 4.6cm$^2$, while the course
(3) and (4) are just above the pain threshold.  
%As a consequence, if we assume a width of SD of about 1cm, corresponding to
%about 15 seconds of hyperexcitabilty during which the neurons get completely
%depleted 

In a nondimensionalized generic model, we can only give an estimate of the
surface area of the pain threshold. In fact, we rely on some self-consistency
which can have biased the result as discussed briefly in the following.

To estimate surface area in units of cm$^2$, the nondimensionalization of the
SD model needs to be considered.  We could have started with bistable kinetics
for the activator, similar to that as suggested in the GHH model for $u=[K^+]_e$ with
three roots at 3mM and 55mM, for the stable fixed points, resting state and
depleted state, respectively, and 10mM for the threshold, a value that is under
physiological conditions well preserved and therefore called the ceiling level of
$[K^+]_e$\cite{HEI77}.  Also the diffusion coefficient for potassium ions in
the brain can be used. This would result in an equation structurally similar to
Eq.~(1) but in a dimensionalized form.  If we were to add linear kinetics for
an immobilized inhibitor (for the sake of argument, set in Eq.~(2) $K=0$),
there is a total of 7 parameters (4 for the cubic rate function of $u$, the
diffusion coefficient, and 2 for the linear rate function of $v$).  We can
scale 4 quantities : the 2 concentrations of $u$ and $v$ as well as time and
space. Therefore, only 3 free parameters remain in a nondimensionalized                                                    
model.  Since the rate function in Eq.~(2) (still with $K=0$) is only a
function of $u$ and not also $v$, this reduces further to the two parameter
$\beta$ and $\varepsilon$---for the weak limit, we consider here, this is
canonical. Weak limit refers to the fact that the threshold of the local
dynamics ($D=0$) is large and the bifurcation that determines the oscillatory
behavior (type I or type II) is irrelevant.

For any set $(\varepsilon, \beta)$ in the parameter plane, we can reconstruct
dimensionalized space and time scales, because we know both the speed of SD and
the width of the wave profile \cite{DAH08}. This procedure can be performed in
a similar way with $K\neq0$, although we select by the choice of the mean field
feedback the size of the critical surface area. This is because the choice of K selects
a specific path in parameter space through the saddle-node bifurcation and
therefore also the ``ghost'' behavior, which is indirectly  linked to the
obtained value of the pain threshold, see Ref.~\cite{DAH12b} for a detailed
discussion. For the values used here, we have approximated the dimensionalized
time and space scales $t_d$ and $x_d$ (and $y_d$) by $t_d=t/10$ and $x_d=x/8$,
where $t$ and $x$ are nondimensionalized time and space scales from
Eqs.~(1)-(2), respectively.

\section{Dynamical disease and neuromodulation} \label{sec:neuromodulation}  

%XXXRAUS?  The migraine generator (Sec.~\ref{sec:mg}) and the spreading
%depression (Sec.~\ref{sec:sd}) theory of migraine are not mutually exclusive
%theories of the cause of episodic migraine. Many evidence speaks to the fact
%that both theories can be validly maintained. It may be that migraine subtypes
%MWoA and MWA have different pathophysiological mechanisms, though this seems
%rather unlikely.  Likewise, it cannot be ruled out that the phenotype of each
%classification subtype has different pathophysiological mechanisms.  
%
%We still let space for either of these possibilities, if we connect the dots
%and combine the network structure of the migraine generator and dynamics of
%spreading depression.  
%
%%, which at a later stage should %include an detailed modelling of the
%dynamics of the MGN, does not

%  The explanatory power of this suggestion is
%twofold: ... can a course of cortical SD stay clinically silent. 

It is often not helpful or even meaningful to talk about an upstream cause for
a composite dynamical system being out of control, if the control is
established in  closed loop feedback between the composites. It is likely that
migraine episodes are caused by events that should be described as
transitions in the dynamical state of the brain. And it can take various
triggers to make this transition but the cause is inherently in the dynamics.
Therefore, let us address the  question: Why do we want to understand migraine
as a dynamical disease?  

``The significance of identifying a dynamical disease is that it should be
possible to develop therapeutic strategies based on our understanding of
dynamics combined with manipulations of the physiological parameters back into
the normal ranges''\cite{BEL95}. This quote provides a general answer to 
this question and to why therapeutic strategies that rely on
the dynamics will be considered in this section.

The focus in this study is on neuromodulation techniques. One reason is
because pharmacological treatment is not suitable for any direct dynamical
treatment at high frequencies because most fast time scales are damped due to
pharmacokinetics, that is, absorption, distribution, and metabolism, before
pharmacodynamics sets in---which of course can act indirectly on fast time
scales.  These issues are bypassed in neuromodulation.  Neuromodulation is also
usually (though not always, see below) spatially confined.  Since dynamical
diseases are caused by transitions in both temporal rhythms and spatio-temporal
patterns, neuromodulation is the natural ansatz for their treatment.  ``The
headache future is bright for neuromodulation techniques, if we manage to
understand how they work.'' A quote from a slide by Jan Schoenen (with 
permission).

Another reason is that various minimally invasive and noninvasive
neuromodulation techniques are available and have been tested clinically in
migraine\cite{MAG12}.  To date, these treatment options are usually only
considered in migraine in chronic cases of several disabling attacks per month
that are refractory to other current treatment. But they might become available
in episodic migraine, if safety issues are appropriately resolved.

\subsection{Phase-dependent mode of action}

\begin{figure} \begin{center}
\includegraphics[width=\columnwidth]{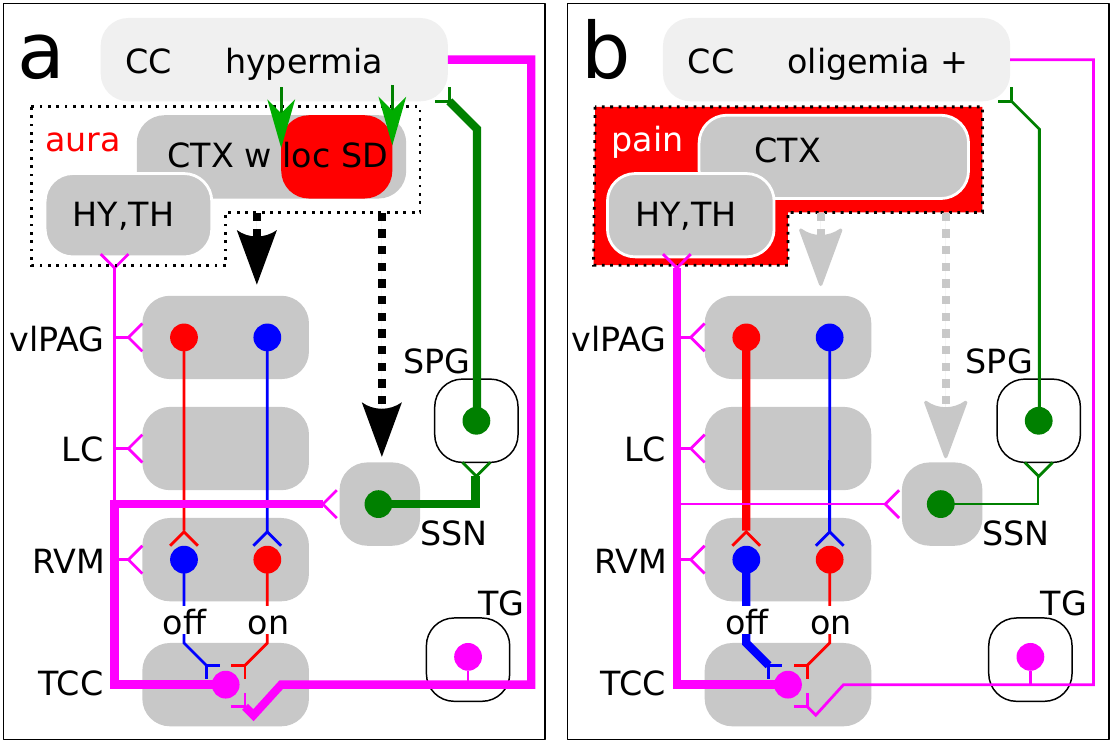}
\caption{\label{fig:painNetworkMigraine} (color online) Suggested facilitated flow of pain traffic in the reduced
migraine generator network. (a) Flow during ignition of SD; (b) after central sensitisation.}
\end{center} \end{figure}

Possible target regions are the brain structures depicted in
Fig.~\ref{fig:painNetwork}. At least two distinct dynamical situations during a
migraine episode should be distinguished because the optimal stimulation
protocol is likely to depend not only on the mode of action but also  on these
different phases. The ignition phase (IP, Fig.~\ref{fig:painNetworkMigraine}a)
where SD is actively controlled by hyperemia, and the longer lasting acute phase
(AP, Fig.~\ref{fig:painNetworkMigraine}b), including oligemia ($\sim2$h) and
sustained pain activity, which can last in migraine from 4 to 72h. 

\begin{figure}[b] \begin{center}
\includegraphics[width=0.75\columnwidth]{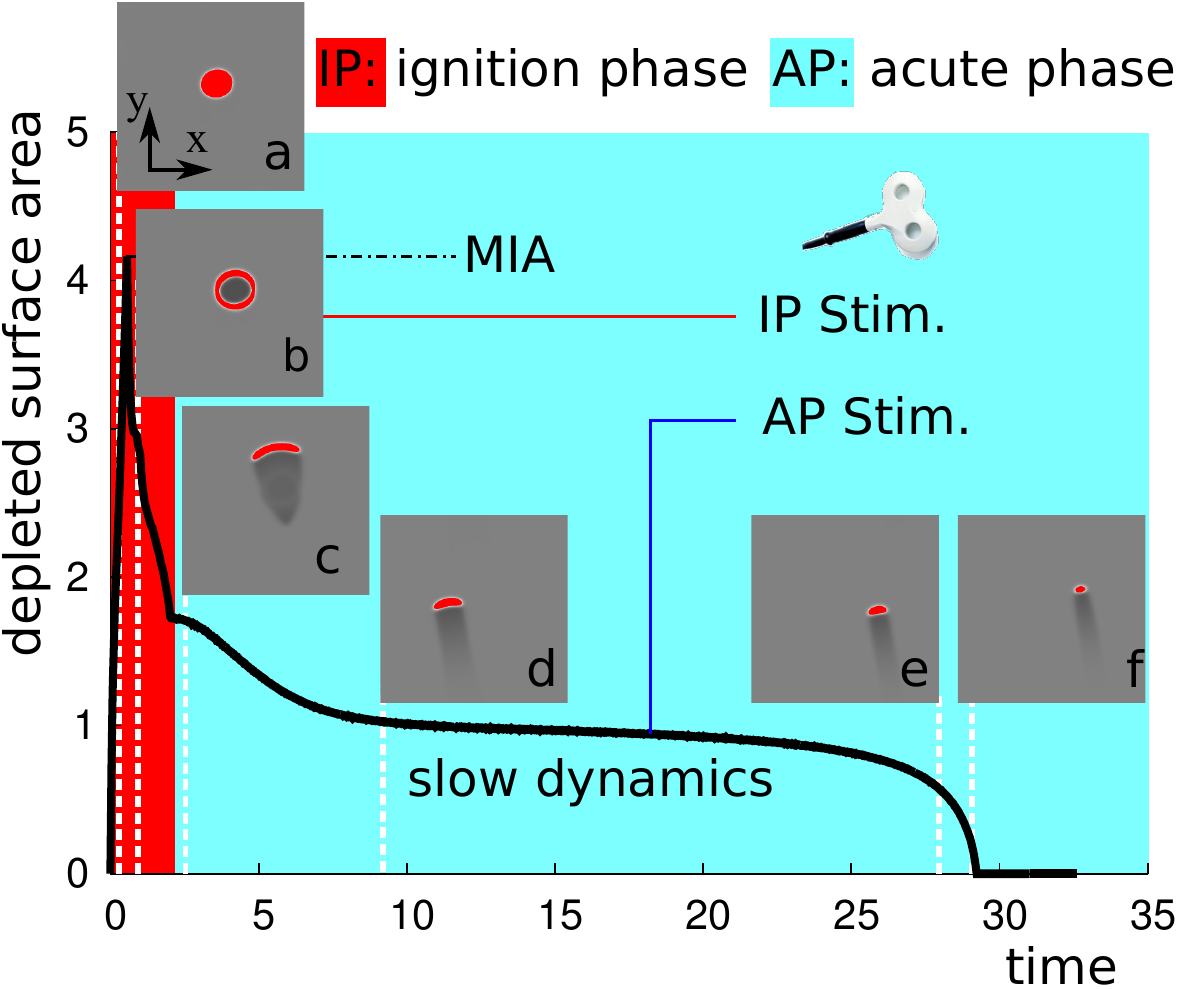}
\caption{\label{fig:sizeVsTime} (color online) Simulated course of cortical
surface area depleted by during SD. Insets: snapshots of simulated SD wave in 2D, top view,
(a) ignition shortly before MIA is reached; (b) ring wave before breaking into
discontineous wave; (c) discontineous wave; (d) propagation near ``ghost'',
note the new reference frame; (e) propagation behind the ghost, new
reference frame; (f) collapse.} \end{center} \end{figure}

The IP lasts about as long as SD circumscribes a large surface area in the
cortical tissue, which is assumed to be above the pain threshold,
Fig.~\ref{fig:sizeVsTime}. During the acute aura phase, there is a cortical
spatio-temporal pattern of both hyperemia and oligemia related to the SD
pattern. This pattern might project even to the nuclei of the MGN, that is, it
is mapped in a topographic representation. The acute pain phase can overlap and
last up to 72h. Therefore Fig.~\ref{fig:painNetworkMigraine} oversimplifies
these phases but gives a first hint on the dominant dynamics. In particular,
Fig.~\ref{fig:painNetworkMigraine}a shows the traffic that eventually results
in central sensitization, while Fig.~\ref{fig:painNetworkMigraine}b shows the
facilitated pain traffic during central sensitization (cf.~Sec.~\ref{sec:mg}).
As a consequence, different neuromodulation targets and/or stimulation
protocols should be developed and applied for IP and AP. Various techniques are
available, as briefly reviewed in the following.

%the migraine stage where the headache episode is full-blown.

\subsection{Invasive neuromodulation}

It is worthwhile to note that patients with chronic pain syndromes---but
headache-free---developed headaches with migrainous features following
implantation of deep brain stimulation (DBS) electrodes that target in
periaqueductal grey (PAG)\cite{VEL03,RAS05}. This was taken as evidence for the
significant role of PAG in the migraine generator (see Sec.~\ref{sec:mg}, cf.
Ref.~\cite{BOR12}).  Although posterior hypothalamic DBS is applied to
cluster headache, DBS as an invasive method does not play a significant role in
headache treatment to date.  The 3\% risk of bleeding is only within the upper
range reported for DBS in movement disorders and other than in these disorders
(cf.~Refs.\cite{SCH10h,RUB12}), the mode of action remains unclear for
headache.  In fact, even placebo has not been conclusively ruled
out\cite{MAG12}.

Therefore, current development efforts are primarily on minimally invasive, and
noninvasive (see below) neuromodulation. In IP, the sphenopalatine ganglion
(SPG, Sec.~II\,D) plays a central role, therefore minimally invasive
sphenopalatine ganglion stimulation might be suited in particular in this
phase. An implantable battery-free SPG stimulator has been developed to
apply on-demand stimulation for the treatment of migraine and other primary
headache\cite{PAE12}.  In a pilot study (n=11), the potential of electrical
stimulation of the SPG has been shown, 2 patients were pain-free within 3
minutes of stimulation, 3 had pain reduction, 5 had no pain relief, 1 was not
stimulated\cite{TEP09}. However, in this pilot study, the headache was allowed
to intensify up to 6 hours before stimulation was initiated, which means that
mode of action is to be considered in the AP paradigm, where SPG probably takes
a less direct influence, see Fig.~\ref{fig:painNetworkMigraine}b.

A promising target is the TCC in both AP and IP. With occipital nerve
stimulation (ONS), the superior cervical spinal cord of the TCC is accessible
with minimally invasive peripheral neuromodulation.  Subcutaneously (beneath
skin) placed electrodes that start at the level of C1 and transverse to C3
affect the corresponding occipital nerve\cite{POP03,SCH07i}. ONS is currently
tested\cite{MAG07,MAG11}.  To develop therapeutic strategies depending on a
specific mode of action in ONS, quantitative approaches for the migraine
generator have to be further developed to understand how the facilitated pain
traffic during central sensitization (Fig.~\ref{fig:painNetworkMigraine}b)
can be influenced via the ophthalmic division of the trigeminal nerve.

\subsection{Noninvasive neuromodulation}

Transcutaneous and transcranial stimulation are the two noninvasive modes.
Transcutaneous electrical nerve stimulation (TENS) modulates neural activity by
targeting noninvasively peripheral nerves (cf.~ONS).  In migraine, a
supraorbital transcutaneous stimulation (STS) was tested with a portable device
that looks like a silver headband and targets the ophthalmic division of the
trigeminal nerve---the first-order neuron that ends in the TCC. The stimulation was
tested with biphasic rectangular AC impulses at 60Hz.  The therapeutic gain
(26\%) is within the range of those reported for other preventive drug and
nondrug antimigraine treatments\cite{SCH13d}. But like with ONS, the mode of
action or even quantitative methods to test paradigms are currently missing,
though it seems suggestive that if central sensitization of second-order
neurons are causing the pain, this method is suppressing not the
sensitivity to noxious stimuli or non-noxious stimuli but the inbound sensory
traffic.

Transcranial stimulation targets the cortex. Two noninvasive neuromodulation
techniques are available in migraine, transcranial magnetic stimulation (TMS)
and  transcranial electrical stimulation (TES).  A carefully sham-controlled
larger trial with a portable TMS devise was undertaken as a promising
noninvasive neurostimulator for disrupting migraine attacks\cite{LIP10}.  The
simplest technology with the least safety issue is probably TES, a method that
therefore might pave the road for noninvasive portable devices for episodic
migraine in less disabled patients in the future.  Two TES versions exist,
transcranial direct current stimulation (tDCS) and alternating current
stimulation (tACS).  The lack of spatial specificity is, however, a critical
point.  Cathodal tDCS inhibits, anodal tDCS increases neuronal firing.
Preliminary evidence was found for patients with chronic migraine having a
positive, but delayed, response to tDCS applied to motor (anodal) and
oribitofrontal (cathode) cortices \cite{DAS12}.  Computational simulations of
current flow through brain regions were used to interpret the effects. With the
help of such simulations, a principal understanding of pain modulation can be
gained, whether this modulation is due to the effect of TES on SD or on the
dynamics of the pain matrix (in chronic pain). Furthermore, the effect of TES
on deeper structures beyond the immediate cortical target regions  can be
estimated, including the cingulate, insula, thalamus, and
brainstem\cite{DAS12}. This is a clinical research model that could serve as a
guide to future investigations.  Effective prophylactic therapy in migraine was
also studied with cathodal tDCS over the visual cortex with the anode overlying
the motor and sensory cortices\cite{ANT11a}.  When alternating current (AC) is
applied over the scalp it is known as tACS, but tACS has not been clinically
tested in migarine.

\subsection{Model-based control}
\label{sec:ModelBasedControl}

The above mentioned techniques need a model of the pathophysiological events in
the target region to optimize the stimulation protocol. This is even mandatory
in a closed loop feedback control, such as a Kalman filter, but it is also
important for open loop control, which is the current mode in all
neuromodulation for migraine.  Simply said, given that we know how the hardware can
interface with the wetware, what is the best software? 

Quantitative models for SD are already developed in great detail. Not only
macroscopic models are available, such as suggested and briefly introduced
here, but also detailed models including the electrophsiological events on a
cellular level, for a review see Ref.~\cite{MIU07}. This opens up a model-based
approach to target SD in the cortex.  

Macroscopic equations of motion for SD, as Eqs.~(1)-(2)---which also are
considered as ``toy models''---have two advantages. First, they allow extensive
computational simulations  of spatio-temporal pattern that extend in the case
of SD in migraine over several centimeters and last up to one hour.  A
microscopic model of SD describing the dynamics on the millisecond scale of
single cells to calculate the noxious signatures that are transmitted into the
meninges during one hour over several square centimeters is to crack a nut with
a sledgehammer.  However, the main advantage of such generic models lies in the
fact that they allow insight in the phase space structure of the whole class of
models they represent,
Fig.~\ref{fig:excitableMediumPhasePortraitMeanField-all}.  In this sense they
are generic and well suited in particular to develop open loop control
stimulation protocols.    

Given the predicted spatio-temporal signatures following from the model in
Eq.~(1)-(2), stimulation protocols in the ignition phase and the acute
phase have different control aims. The acute phase can be further subdivided
into one with the aura present and one with only the headache. Furthermore, the
prodromal phase exists. 

The most obvious stimulation protocol can be derived for the acute phase with
the aura present. In this phase the slow dynamics are governed by ghost
behavior after a saddle-node bifurcation,
Fig.~\ref{fig:excitableMediumPhasePortraitMeanField-all}d.  Noise can speed up
the transition time in this case and shorten the aura. Although this is not the
major contribution to the pain phase as SD is below the pain threshold, once
this threshold is crossed during ignition, it is likely beneficial to
suppress any further release of noxious or inflammatory mediators by the decaying SD wave. 

According to our prediction, the most critical phase is the ignition phase.  In
this phase the pain threshold will be usually crossed. It is also the shortest
lasting phase of only a couple of  minutes. This phase is therefore likely to
be missed, if neuromodulation is not used in a closed-loop control with
some sort of feedback. 

Therefore, the prodromal phase becomes in particular valuable for prophylactic
protocols.  In this phase, cortical excitability might be temporally
decreased to lower the upcoming ignition.  Prophylactic protocols with cathodal
transcranial direct current stimulation of the visual cortex are currently
tested\cite{ANT11a}. Similar approaches should be tested with TMS, which can be
more spatially selective to interact with cortical hot spots as targets.

% these pattern forming mechanism. While TES may seem
%promising its lack of spatial specificity might cause various side effects.
%The following stimulation protocol takes into account the critical phases in
%episodic migraine and should be tested after a cohort imaging  study as tested
%the predicted correlations of the spatio-temporal signatures of SD and the
%migraine subtypes: In the ignition phase, an inhibitory stimulation protocol
%can confine the initial spatial nucleation, while in the acute aura phase
%noise protocol 
%
%
  
In the acute pain phase, SD is not a target anymore. Therefore the model
contributes nothing to this phase, except, as already noted above, that
according to the proposed mechanism transcutaneous stimulation will in this
case target the inbound sensory traffic.

\section{Summary} \label{sec:discussion}

The migraine generator (Sec.~\ref{sec:mg}) and the spreading depression
(Sec.~\ref{sec:sd}) theory of migraine are not mutually exclusive theories of
the cause of episodic migraine. Sufficient evidence speaks to the fact that both
theories can be validly maintained. It may be that the migraine subtypes
MWoA and MWA have different pathophysiological mechanisms, though this seems
rather unlikely.  Likewise, it cannot be ruled out that the phenotype of each
classification subtype has different pathophysiological mechanisms.  

The proposed mechanism still allows space for either of these possibilities.  The
unified framework, however, does not rely on either of these options.  We
propose that stereotype spatio-temporal signatures of SD
(Fig.~\ref{fig:bestiaMIAvsTAA})  can selectively drive rhythms in the MGN,
which in turn is part of a larger network (Fig.~\ref{fig:painNetwork}) that
controls the spatio-temporal dynamics of SD in the first place. 

The question of the most upstream events in episodic migraine lead to a long
standing controversy. To my mind, the settlement of this controversy points strongly
to the need to complement clinical and experimental data with quantitative
approaches, in particular computational simulations and mathematical analysis
of the dynamical interaction between the migraine generator and spreading
depression. This study can only outline a framework for such quantitative
approaches.

\section*{Acknowledgments} The author kindly acknowledges the support from the
Bundesministerium f\"ur Bildung und Forschung (BMBF 01GQ1109). This research
has been supported in part also by the Mathematical Biosciences Institute at
the Ohio State University and the National Science Foundation under Grant
No. DMS 0931642.

\section*{References}


\begin{thebibliography}{10}%
\makeatletter
\providecommand \@ifxundefined [1]{%
 \ifx #1\undefined \expandafter \@firstoftwo
 \else \expandafter \@secondoftwo
\fi
}%
\providecommand \@ifnum [1]{%
 \ifnum #1\expandafter \@firstoftwo
 \else \expandafter \@secondoftwo
\fi
}%
\providecommand \enquote [1]{``#1''}%
\providecommand \bibnamefont  [1]{#1}%
\providecommand \bibfnamefont [1]{#1}%
\providecommand \citenamefont [1]{#1}%
\providecommand\href[0]{\@sanitize\@href}%
\providecommand\@href[1]{\endgroup\@@startlink{#1}\endgroup\@@href}%
\providecommand\@@href[1]{#1\@@endlink}%
\providecommand \@sanitize [0]{\begingroup\catcode`\&12\catcode`\#12\relax}%
\@ifxundefined \pdfoutput {\@firstoftwo}{%
 \@ifnum{\z@=\pdfoutput}{\@firstoftwo}{\@secondoftwo}%
}{%
 \providecommand\@@startlink[1]{\leavevmode}%
 \providecommand\@@endlink[0]{}%
}{%
 \providecommand\@@startlink[1]{%
  \leavevmode
  \pdfstartlink
   attr{/Border[0 0 1 ]/H/I/C[0 1 1]}%
   user{/Subtype/Link/A<</Type/Action/S/URI/URI(#1)>>}%
  \relax
 }%
 \providecommand\@@endlink[0]{\pdfendlink}%
}%
\providecommand \url  [0]{\begingroup\@sanitize \@url }%
\providecommand \@url [1]{\endgroup\@href {#1}{\urlprefix}}%
\providecommand \urlprefix [0]{URL }%
\providecommand \Eprint[0]{\href }%
\@ifxundefined \urlstyle {%
  \providecommand \doi [1]{doi:\discretionary{}{}{}#1}%
}{%
  \providecommand \doi [0]{doi:\discretionary{}{}{}\begingroup
  \urlstyle{rm}\Url }%
}%
\providecommand \doibase [0]{http://dx.doi.org/}%
\providecommand \Doi[1]{\href{\doibase#1}}%
\providecommand \selectlanguage [0]{\@gobble}%
\providecommand \bibinfo [0]{\@secondoftwo}%
\providecommand \bibfield [0]{\@secondoftwo}%
\providecommand \translation [1]{[#1]}%
\providecommand \BibitemOpen[0]{}%
\providecommand \bibitemStop [0]{}%
\providecommand \bibitemNoStop [0]{.\EOS\space}%
\providecommand \EOS [0]{\spacefactor3000\relax}%
\providecommand \BibitemShut [1]{\csname bibitem#1\endcsname}%
%</preamble>
\bibitem{STO07}%
  \BibitemOpen
  \bibfield{author}{%
  \bibinfo {author} {\bibfnamefont{L.J.}~\bibnamefont{Stovner}}, \bibinfo
  {author} {\bibfnamefont{K.}~\bibnamefont{Hagen}}, \bibinfo {author}
  {\bibfnamefont{R.}~\bibnamefont{Jensen}}, \bibinfo {author}
  {\bibfnamefont{Z.}~\bibnamefont{Katsarava}}, \bibinfo {author}
  {\bibfnamefont{R.B.}~\bibnamefont{Lipton}}, \bibinfo {author}
  {\bibfnamefont{A.I.}~\bibnamefont{Scher}}, \bibinfo {author}
  {\bibfnamefont{T.J.}~\bibnamefont{Steiner}},\ and\ \bibinfo {author}
  {\bibfnamefont{J.A.}\ \bibnamefont{Zwart}},\ }%
  \bibfield{title}{%
  \enquote{\bibinfo {title} {The global burden of headache: a documentation of
  headache prevalence and disability worldwide},}\ }%
  \bibfield{journal}{%
  \bibinfo {journal} {Cephalalgia}\ }%
  \textbf{\bibinfo {volume} {27}},\ \bibinfo {pages} {193--210} (\bibinfo
  {year} {2007})\BibitemShut{NoStop}%
\bibitem{HAN12a}%
  \BibitemOpen
  \bibfield{author}{%
  \bibinfo {author} {\bibfnamefont{J.~M.}\ \bibnamefont{Hansen}}, \bibinfo
  {author} {\bibfnamefont{R.~B.}\ \bibnamefont{Lipton}}, \bibinfo {author}
  {\bibfnamefont{D.~W.}\ \bibnamefont{Dodick}}, \bibinfo {author}
  {\bibfnamefont{S.~D.}\ \bibnamefont{Silberstein}}, \bibinfo {author}
  {\bibfnamefont{J.~R.}\ \bibnamefont{Saper}}, \bibinfo {author}
  {\bibfnamefont{S.~K.}\ \bibnamefont{Aurora}}, \bibinfo {author}
  {\bibfnamefont{P.~J.}\ \bibnamefont{Goadsby}},\ and\ \bibinfo {author}
  {\bibfnamefont{A.}~\bibnamefont{Charles}},\ }%
  \bibfield{title}{%
  \enquote{\bibinfo {title} {{{M}igraine headache is present in the aura phase:
  a prospective study}},}\ }%
  \bibfield{journal}{%
  \bibinfo {journal} {Neurology}\ }%
  \textbf{\bibinfo {volume} {79}},\ \bibinfo {pages} {2044--2049} (\bibinfo
  {year} {2012})\BibitemShut{NoStop}%
\bibitem{MAC87}%
  \BibitemOpen
  \bibfield{author}{%
  \bibinfo {author} {\bibfnamefont{M.C.}\ \bibnamefont{Mackey}}\ and\ \bibinfo
  {author} {\bibfnamefont{J.G.}\ \bibnamefont{Milton}},\ }%
  \bibfield{title}{%
  \enquote{\bibinfo {title} {{{D}ynamical diseases}},}\ }%
  \bibfield{journal}{%
  \bibinfo {journal} {Ann. N. Y. Acad. Sci.}\ }%
  \textbf{\bibinfo {volume} {504}},\ \bibinfo {pages} {16--32} (\bibinfo {year}
  {1987})\BibitemShut{NoStop}%
\bibitem{WEI95a}%
  \BibitemOpen
  \bibfield{author}{%
  \bibinfo {author} {\bibfnamefont{C.}~\bibnamefont{Weiller}}, \bibinfo
  {author} {\bibfnamefont{A.}~\bibnamefont{May}}, \bibinfo {author}
  {\bibfnamefont{V.}~\bibnamefont{Limmroth}}, \bibinfo {author}
  {\bibfnamefont{M.}~\bibnamefont{Juptner}}, \bibinfo {author}
  {\bibfnamefont{H.}~\bibnamefont{Kaube}}, \bibinfo {author}
  {\bibfnamefont{R.V.}\ \bibnamefont{Schayck}}, \bibinfo {author}
  {\bibfnamefont{H.H.}\ \bibnamefont{Coenen}},\ and\ \bibinfo {author}
  {\bibfnamefont{H.C.}\ \bibnamefont{Diener}},\ }%
  \bibfield{title}{%
  \enquote{\bibinfo {title} {{{B}rain stem activation in spontaneous human
  migraine attacks}},}\ }%
  \bibfield{journal}{%
  \bibinfo {journal} {Nat. Med.}\ }%
  \textbf{\bibinfo {volume} {1}},\ \bibinfo {pages} {658--660} (\bibinfo {year}
  {1995})\BibitemShut{NoStop}%
\bibitem{WEL01}%
  \BibitemOpen
  \bibfield{author}{%
  \bibinfo {author} {\bibfnamefont{KMA}\ \bibnamefont{Welch}}, \bibinfo
  {author} {\bibfnamefont{V.}~\bibnamefont{Nagesh}}, \bibinfo {author}
  {\bibfnamefont{S.K.}~\bibnamefont{Aurora}},\ and\ \bibinfo {author}
  {\bibfnamefont{N.}~\bibnamefont{Gelman}},\ }%
  \bibfield{title}{%
  \enquote{\bibinfo {title} {Periaqueductal gray matter dysfunction in
  migraine: cause or the burden of illness?}.}\ }%
  \bibfield{journal}{%
  \bibinfo {journal} {Headache: The Journal of Head and Face Pain}\ }%
  \textbf{\bibinfo {volume} {41}},\ \bibinfo {pages} {629--637} (\bibinfo
  {year} {2001})\BibitemShut{NoStop}%
\bibitem{FOX05}%
  \BibitemOpen
  \bibfield{author}{%
  \bibinfo {author} {\bibfnamefont{Anthony~W}\ \bibnamefont{Fox}},\ }%
  \bibfield{title}{%
  \enquote{\bibinfo {title} {Time-series data and the ``migraine
  generator''},}\ }%
  \bibfield{journal}{%
  \bibinfo {journal} {Headache}\ }%
  \textbf{\bibinfo {volume} {45}},\ \bibinfo {pages} {920--925} (\bibinfo
  {year} {2005})\BibitemShut{NoStop}%
\bibitem{BOR12}%
  \BibitemOpen
  \bibfield{author}{%
  \bibinfo {author} {\bibfnamefont{D}~\bibnamefont{Borsook}}\ and\ \bibinfo
  {author} {\bibfnamefont{R}~\bibnamefont{Burstein}},\ }%
  \bibfield{title}{%
  \enquote{\bibinfo {title} {The enigma of the dorsolateral pons as a migraine
  generator},}\ }%
  \bibfield{journal}{%
  \bibinfo {journal} {Cephalalgia}\ }%
  \textbf{\bibinfo {volume} {32}},\ \bibinfo {pages} {803--812} (\bibinfo
  {year} {2012})\BibitemShut{NoStop}%
\bibitem{BUR05a}%
  \BibitemOpen
  \bibfield{author}{%
  \bibinfo {author} {\bibfnamefont{R.}~\bibnamefont{Burstein}}\ and\
  \bibinfo {author} {\bibfnamefont{M.}~\bibnamefont{Jakubowski}},\ }%
  \bibfield{title}{%
  \enquote{\bibinfo {title} {Unitary hypothesis for multiple triggers of the
  pain and strain of migraine},}\ }%
  \bibfield{journal}{%
  \bibinfo {journal} {J.~Comp. Neurology}\ }%
  \textbf{\bibinfo {volume} {493}},\ \bibinfo {pages} {9--14} (\bibinfo {year}
  {2005})\BibitemShut{NoStop}%
\bibitem{HEi09a}%
  \BibitemOpen
  \bibfield{author}{%
  \bibinfo {author} {\bibfnamefont{M.~M.}\ \bibnamefont{Heinricher}}, \bibinfo
  {author} {\bibfnamefont{I.}~\bibnamefont{Tavares}}, \bibinfo {author}
  {\bibfnamefont{J.~L.}\ \bibnamefont{Leith}},\ and\ \bibinfo {author}
  {\bibfnamefont{B.~M.}\ \bibnamefont{Lumb}},\ }%
  \bibfield{title}{%
  \enquote{\bibinfo {title} {{{D}escending control of nociception:
  {S}pecificity, recruitment and plasticity}},}\ }%
  \bibfield{journal}{%
  \bibinfo {journal} {Brain Res Rev}\ }%
  \textbf{\bibinfo {volume} {60}},\ \bibinfo {pages} {214--225} (\bibinfo
  {year} {2009})\BibitemShut{NoStop}%
\bibitem{IAN10}%
  \BibitemOpen
  \bibfield{author}{%
  \bibinfo {author} {\bibfnamefont{GD}~\bibnamefont{Iannetti}}\ and\ \bibinfo
  {author} {\bibfnamefont{A}~\bibnamefont{Mouraux}},\ }%
  \bibfield{title}{%
  \enquote{\bibinfo {title} {From the neuromatrix to the pain matrix (and
  back)},}\ }%
  \bibfield{journal}{%
  \bibinfo {journal} {Exp. Brain Res.}\ }%
  \textbf{\bibinfo {volume} {205}},\ \bibinfo {pages} {1--12} (\bibinfo {year}
  {2010})\BibitemShut{NoStop}%
\bibitem{PUR78}%
  \BibitemOpen
  \bibfield{author}{%
  \bibinfo {author} {\bibfnamefont{M.J.}~\bibnamefont{Purves}},\ }%
  \bibfield{title}{%
  \enquote{\bibinfo {title} {Do vasomotor nerves significantly regulate
  cerebral blood flow?}.}\ }%
  \bibfield{journal}{%
  \bibinfo {journal} {Circulation Research}\ }%
  \textbf{\bibinfo {volume} {43}},\ \bibinfo {pages} {485--493} (\bibinfo
  {year} {1978})\BibitemShut{NoStop}%
\bibitem{PRI06}%
  \BibitemOpen
  \bibfield{author}{%
  \bibinfo {author} {\bibfnamefont{Joseph~L}\ \bibnamefont{Price}},\ }%
  \bibfield{title}{%
  \enquote{\bibinfo {title} {Prefrontal cortical networks related to visceral
  function and mood},}\ }%
  \bibfield{journal}{%
  \bibinfo {journal} {Annals of the New York Academy of Sciences}\ }%
  \textbf{\bibinfo {volume} {877}},\ \bibinfo {pages} {383--396} (\bibinfo
  {year} {2006})\BibitemShut{NoStop}%
\bibitem{KEA01}%
  \BibitemOpen
  \bibfield{author}{%
  \bibinfo {author} {\bibfnamefont{Kevin~A}\ \bibnamefont{Keay}}\ and\ \bibinfo
  {author} {\bibfnamefont{Richard}\ \bibnamefont{Bandler}},\ }%
  \bibfield{title}{%
  \enquote{\bibinfo {title} {Parallel circuits mediating distinct emotional
  coping reactions to different types of stress},}\ }%
  \bibfield{journal}{%
  \bibinfo {journal} {Neuroscience \& Biobehavioral Reviews}\ }%
  \textbf{\bibinfo {volume} {25}},\ \bibinfo {pages} {669--678} (\bibinfo
  {year} {2001})\BibitemShut{NoStop}%
\bibitem{AKE11}%
  \BibitemOpen
  \bibfield{author}{%
  \bibinfo {author} {\bibfnamefont{S.}~\bibnamefont{Akerman}}, \bibinfo
  {author} {\bibfnamefont{P.R.}\ \bibnamefont{Holland}},\ and\ \bibinfo
  {author} {\bibfnamefont{P.J.}\ \bibnamefont{Goadsby}},\ }%
  \bibfield{title}{%
  \enquote{\bibinfo {title} {{{D}iencephalic and brainstem mechanisms in
  migraine}},}\ }%
  \bibfield{journal}{%
  \bibinfo {journal} {Nat. Rev. Neurosci.}\ }%
  \textbf{\bibinfo {volume} {12}},\ \bibinfo {pages} {570--584} (\bibinfo
  {year} {2011})\BibitemShut{NoStop}%
\bibitem{MEN11a}%
  \BibitemOpen
  \bibfield{author}{%
  \bibinfo {author} {\bibfnamefont{L.~M.}\ \bibnamefont{Mendell}},\ }%
  \bibfield{title}{%
  \enquote{\bibinfo {title} {{{C}omputational functions of neurons and circuits
  signaling injury: relationship to pain behavior}},}\ }%
  \bibfield{journal}{%
  \bibinfo {journal} {Proc. Natl. Acad. Sci. U.S.A.}\ }%
  \textbf{\bibinfo {volume} {108 Suppl 3}},\ \bibinfo {pages} {15596--15601}
  (\bibinfo {year} {2011})\BibitemShut{NoStop}%
\bibitem{BUR00}%
  \BibitemOpen
  \bibfield{author}{%
  \bibinfo {author} {\bibfnamefont{R.}~\bibnamefont{Burstein}}, \bibinfo
  {author} {\bibfnamefont{M.F.}\ \bibnamefont{Cutrer}},\ and\ \bibinfo
  {author} {\bibfnamefont{D.}~\bibnamefont{Yarnitsky}},\ }%
  \bibfield{title}{%
  \enquote{\bibinfo {title} {{{T}he development of cutaneous allodynia during a
  migraine attack clinical evidence for the sequential recruitment of spinal
  and supraspinal nociceptive neurons in migraine}},}\ }%
  \bibfield{journal}{%
  \bibinfo {journal} {Brain}\ }%
  \textbf{\bibinfo {volume} {123}},\ \bibinfo {pages} {1703--1709} (\bibinfo
  {year} {2000})\BibitemShut{NoStop}%
\bibitem{BRI89}%
  \BibitemOpen
  \bibfield{author}{%
  \bibinfo {author} {\bibfnamefont{N.F.}~\bibnamefont{Britton}}\ and\ \bibinfo
  {author} {\bibfnamefont{S.M.}~\bibnamefont{Skevington}},\ }%
  \bibfield{title}{%
  \enquote{\bibinfo {title} {A mathematical model of the gate control theory of
  pain},}\ }%
  \bibfield{journal}{%
  \bibinfo {journal} {Journal of theoretical biology}\ }%
  \textbf{\bibinfo {volume} {137}},\ \bibinfo {pages} {91--105} (\bibinfo
  {year} {1989})\BibitemShut{NoStop}%
\bibitem{DRE11}%
  \BibitemOpen
  \bibfield{author}{%
  \bibinfo {author} {\bibfnamefont{J.P.}\ \bibnamefont{Dreier}},\ }%
  \bibfield{title}{%
  \enquote{\bibinfo {title} {{{T}he role of spreading depression, spreading
  depolarization and spreading ischemia in neurological disease}},}\ }%
  \bibfield{journal}{%
  \bibinfo {journal} {Nat. Med.}\ }%
  \textbf{\bibinfo {volume} {17}},\ \bibinfo {pages} {439--447} (\bibinfo
  {year} {2011})\BibitemShut{NoStop}%
\bibitem{SOM01}%
  \BibitemOpen
  \bibfield{author}{%
  \bibinfo {author} {\bibfnamefont{G.~G.}\ \bibnamefont{Somjen}},\ }%
  \bibfield{title}{%
  \enquote{\bibinfo {title} {Mechanisms of spreading depression and hypoxic
  spreading depression-like depolarization},}\ }%
  \bibfield{journal}{%
  \bibinfo {journal} {Physiol. Rev.}\ }%
  \textbf{\bibinfo {volume} {81}},\ \bibinfo {pages} {1065--1096} (\bibinfo
  {year} {2001})\BibitemShut{NoStop}%
\bibitem{DRE12}%
  \BibitemOpen
  \bibfield{author}{%
  \bibinfo {author} {\bibfnamefont{J.P.}\ \bibnamefont{Dreier}}, \bibinfo
  {author} {\bibfnamefont{T.M.}\ \bibnamefont{Isele}}, \bibinfo {author}
  {\bibfnamefont{C.}~\bibnamefont{Reiffurth}}, \bibinfo {author}
  {\bibfnamefont{S.A.}\ \bibnamefont{Kirov}}, \bibinfo {author}
  {\bibfnamefont{M.A.}\ \bibnamefont{Dahlem}},\ and\ \bibinfo {author}
  {\bibfnamefont{O.}~\bibnamefont{Herreras}},\ }%
  \bibfield{title}{%
  \enquote{\bibinfo {title} {Is spreading depolarization characterized by an
  abrupt, massive release of gibbs free energy from the human brain cortex?}.}\
  }%
  \bibfield{journal}{%
  \Doi{10.1177/1073858412453340}{\bibinfo {journal} {Neuroscientist}}\ }%
  \textbf{\bibinfo {volume} {19}},\ \bibinfo {pages} {25--42} (\bibinfo {year}
  {2013})\BibitemShut{NoStop}%
\bibitem{MOS93}%
  \BibitemOpen
  \bibfield{author}{%
  \bibinfo {author} {\bibfnamefont{E.}~\bibnamefont{Mosekilde}}, \bibinfo
  {author} {\bibfnamefont{J.S.}\ \bibnamefont{Thomson}}, \bibinfo {author}
  {\bibfnamefont{C.}~\bibnamefont{Knudsen}},\ and\ \bibinfo {author}
  {\bibfnamefont{R.}~\bibnamefont{Feldberg}},\ }%
  \bibfield{title}{%
  \enquote{\bibinfo {title} {Phase diagrams for periodically driven
  {G}unn-diodes},}\ }%
  \bibfield{journal}{%
  \bibinfo {journal} {Physica~D}\ }%
  \textbf{\bibinfo {volume} {66}},\ \bibinfo {pages} {143} (\bibinfo {year}
  {1993})\BibitemShut{NoStop}%
\bibitem{BOL02}%
  \BibitemOpen
  \bibfield{author}{%
  \bibinfo {author} {\bibfnamefont{H.}~\bibnamefont{Bolay}}, \bibinfo {author}
  {\bibfnamefont{U.}~\bibnamefont{Reuter}}, \bibinfo {author}
  {\bibfnamefont{A.K.}\ \bibnamefont{Dunn}}, \bibinfo {author}
  {\bibfnamefont{Z.}~\bibnamefont{Huang}}, \bibinfo {author}
  {\bibfnamefont{D.A.}\ \bibnamefont{Boas}},\ and\ \bibinfo {author}
  {\bibfnamefont{M.A.}~\bibnamefont{Moskowitz}},\ }%
  \bibfield{title}{%
  \enquote{\bibinfo {title} {{{I}ntrinsic brain activity triggers trigeminal
  meningeal afferents in a migraine model}},}\ }%
  \bibfield{journal}{%
  \bibinfo {journal} {Nat. Med.}\ }%
  \textbf{\bibinfo {volume} {8}},\ \bibinfo {pages} {136--142} (\bibinfo {year}
  {2002})\BibitemShut{NoStop}%
\bibitem{EIK08}%
  \BibitemOpen
  \bibfield{author}{%
  \bibinfo {author} {\bibfnamefont{K.}~\bibnamefont{Eikermann-Haerter}}\ and\
  \bibinfo {author} {\bibfnamefont{M.A.}\ \bibnamefont{Moskowitz}},\ }%
  \bibfield{title}{%
  \enquote{\bibinfo {title} {{{A}nimal models of migraine headache and
  aura}},}\ }%
  \bibfield{journal}{%
  \bibinfo {journal} {Curr. Opin. Neurol.}\ }%
  \textbf{\bibinfo {volume} {21}},\ \bibinfo {pages} {294--300} (\bibinfo
  {year} {2008})\BibitemShut{NoStop}%
\bibitem{TFE11}%
  \BibitemOpen
  \bibfield{author}{%
  \bibinfo {author} {\bibfnamefont{P.~C.}\ \bibnamefont{Tfelt-Hansen}},\ }%
  \bibfield{title}{%
  \enquote{\bibinfo {title} {{{P}ermeability of dura mater: a possible link
  between cortical spreading depression and migraine pain? {A} comment}},}\ }%
  \bibfield{journal}{%
  \bibinfo {journal} {J.~Headache Pain}\ }%
  \textbf{\bibinfo {volume} {12}},\ \bibinfo {pages} {3--4} (\bibinfo {year}
  {2011})\BibitemShut{NoStop}%
\bibitem{ING97}%
  \BibitemOpen
  \bibfield{author}{%
  \bibinfo {author} {\bibfnamefont{B.~K.}\ \bibnamefont{Ingvardsen}}, \bibinfo
  {author} {\bibfnamefont{H.}~\bibnamefont{Laursen}}, \bibinfo {author}
  {\bibfnamefont{U.~B.}\ \bibnamefont{Olsen}},\ and\ \bibinfo {author}
  {\bibfnamefont{A.~J.}\ \bibnamefont{Hansen}},\ }%
  \bibfield{title}{%
  \enquote{\bibinfo {title} {{{P}ossible mechanism of c-fos expression in
  trigeminal nucleus caudalis following cortical spreading depression}},}\ }%
  \bibfield{journal}{%
  \bibinfo {journal} {Pain}\ }%
  \textbf{\bibinfo {volume} {72}},\ \bibinfo {pages} {407--415} (\bibinfo
  {year} {1997})\BibitemShut{NoStop}%
\bibitem{MOS98}%
  \BibitemOpen
  \bibfield{author}{%
  \bibinfo {author} {\bibfnamefont{M.~A.}\ \bibnamefont{Moskowitz}}\ and\
  \bibinfo {author} {\bibfnamefont{R.}~\bibnamefont{Kraig}},\ }%
  \bibfield{title}{%
  \enquote{\bibinfo {title} {{{C}omment on {I}ngvardsen et al., {P}{A}{I}{N},
  72 (1997) 407-415}},}\ }%
  \bibfield{journal}{%
  \bibinfo {journal} {Pain}\ }%
  \textbf{\bibinfo {volume} {76}},\ \bibinfo {pages} {265--267} (\bibinfo
  {year} {1998})\BibitemShut{NoStop}%
\bibitem{KAR13}%
  \BibitemOpen
  \bibfield{author}{%
  \bibinfo {author} {\bibfnamefont{H.}~\bibnamefont{Karatas}}, \bibinfo
  {author} {\bibfnamefont{S.E.}\ \bibnamefont{Erdener}}, \bibinfo {author}
  {\bibfnamefont{Y.}~\bibnamefont{Gursoy-Ozdemir}}, \bibinfo {author}
  {\bibfnamefont{S.}~\bibnamefont{Lule}}, \bibinfo {author}
  {\bibfnamefont{E.}~\bibnamefont{Eren-Kocak}}, \bibinfo {author}
  {\bibfnamefont{Z.~D.}\ \bibnamefont{Sen}},\ and\ \bibinfo {author}
  {\bibfnamefont{T.}~\bibnamefont{Dalkara}},\ }%
  \bibfield{title}{%
  \enquote{\bibinfo {title} {{{S}preading depression triggers headache by
  activating neuronal {P}anx1 channels}},}\ }%
  \bibfield{journal}{%
  \bibinfo {journal} {Science}\ }%
  \textbf{\bibinfo {volume} {339}},\ \bibinfo {pages} {1092--1095} (\bibinfo
  {year} {2013})\BibitemShut{NoStop}%
\bibitem{AYA10}%
  \BibitemOpen
  \bibfield{author}{%
  \bibinfo {author} {\bibfnamefont{Cenk}\ \bibnamefont{Ayata}},\ }%
  \bibfield{title}{%
  \enquote{\bibinfo {title} {Cortical spreading depression triggers migraine
  attack: Pro},}\ }%
  \bibfield{journal}{%
  \Doi{10.1111/j.1526-4610.2010.01647.x}{\bibinfo {journal} {Headache: The
  Journal of Head and Face Pain}}\ }%
  \textbf{\bibinfo {volume} {50}},\ \bibinfo {pages} {725--730} (\bibinfo
  {year} {2010}),\ ISSN \bibinfo {issn} {1526-4610}\BibitemShut{NoStop}%
\bibitem{OLE81}%
  \BibitemOpen
  \bibfield{author}{%
  \bibinfo {author} {\bibfnamefont{J.}~\bibnamefont{Olesen}}, \bibinfo {author}
  {\bibfnamefont{B.}~\bibnamefont{Larsen}},\ and\ \bibinfo {author}
  {\bibfnamefont{M.}~\bibnamefont{Lauritzen}},\ }%
  \bibfield{title}{%
  \enquote{\bibinfo {title} {{{F}ocal hyperemia followed by spreading oligemia
  and impaired activation of r{C}{B}{F} in classic migraine}},}\ }%
  \bibfield{journal}{%
  \bibinfo {journal} {Ann. Neurol.}\ }%
  \textbf{\bibinfo {volume} {9}},\ \bibinfo {pages} {344--352} (\bibinfo {year}
  {1981})\BibitemShut{NoStop}%
\bibitem{LAU87}%
  \BibitemOpen
  \bibfield{author}{%
  \bibinfo {author} {\bibfnamefont{Martin}\ \bibnamefont{Lauritzen}},\ }%
  \bibfield{title}{%
  \enquote{\bibinfo {title} {Cortical spreading depression as a putative
  migraine mechanism},}\ }%
  \bibfield{journal}{%
  \bibinfo {journal} {Trends Neurosci.}\ }%
  \textbf{\bibinfo {volume} {10}},\ \bibinfo {pages} {8--13} (\bibinfo {year}
  {1987}),\ ISSN \bibinfo {issn} {0166-2236}\BibitemShut{NoStop}%
\bibitem{HAD01}%
  \BibitemOpen
  \bibfield{author}{%
  \bibinfo {author} {\bibfnamefont{N.}~\bibnamefont{Hadjikhani}}, \bibinfo
  {author} {\bibfnamefont{M.}~\bibnamefont{Sanchez Del.~Rio.}}, \bibinfo
  {author} {\bibfnamefont{O.}~\bibnamefont{Wu}}, \bibinfo {author}
  {\bibfnamefont{D.}~\bibnamefont{Schwartz}}, \bibinfo {author}
  {\bibfnamefont{D.}~\bibnamefont{Bakker}}, \bibinfo {author}
  {\bibfnamefont{B.}~\bibnamefont{Fischl}}, \bibinfo {author}
  {\bibfnamefont{K.K.}\ \bibnamefont{Kwong}}, \bibinfo {author}
  {\bibfnamefont{F.M.}\ \bibnamefont{Cutrer}}, \bibinfo {author}
  {\bibfnamefont{B.R.}\ \bibnamefont{Rosen}}, \bibinfo {author}
  {\bibfnamefont{R.B.}\ \bibnamefont{Tootell}}, \bibinfo {author}
  {\bibfnamefont{A.G.}\ \bibnamefont{Sorensen}},\ and\ \bibinfo {author}
  {\bibfnamefont{M.A.}\ \bibnamefont{Moskowitz}},\ }%
  \bibfield{title}{%
  \enquote{\bibinfo {title} {{Mechanisms of migraine aura revealed by
  functional MRI in human visual cortex}},}\ }%
  \bibfield{journal}{%
  \bibinfo {journal} {Proc. Natl. Acad. Sci. U.S.A.}\ }%
  \textbf{\bibinfo {volume} {98}},\ \bibinfo {pages} {4687--4692} (\bibinfo
  {year} {2001})\BibitemShut{NoStop}%
\bibitem{LEA45}%
  \BibitemOpen
  \bibfield{author}{%
  \bibinfo {author} {\bibfnamefont{A.A.P.}\ \bibnamefont{Le{\~a}o}}\ and\
  \bibinfo {author} {\bibfnamefont{R.S.}\ \bibnamefont{Morison}},\ }%
  \bibfield{title}{%
  \enquote{\bibinfo {title} {Propagation of cortical spreading depression},}\
  }%
  \bibfield{journal}{%
  \bibinfo {journal} {J. Neurophysiol.}\ }%
  \textbf{\bibinfo {volume} {1}},\ \bibinfo {pages} {33--45} (\bibinfo {year}
  {1945})\BibitemShut{NoStop}%
\bibitem{WOO94}%
  \BibitemOpen
  \bibfield{author}{%
  \bibinfo {author} {\bibfnamefont{R.~P.}\ \bibnamefont{Woods}}, \bibinfo
  {author} {\bibfnamefont{M.}~\bibnamefont{Iacoboni}},\ and\ \bibinfo {author}
  {\bibfnamefont{J.~C.}\ \bibnamefont{Mazziotta}},\ }%
  \bibfield{title}{%
  \enquote{\bibinfo {title} {{{B}rief report: bilateral spreading cerebral
  hypoperfusion during spontaneous migraine headache}},}\ }%
  \bibfield{journal}{%
  \bibinfo {journal} {N.~Engl. J.~Med.}\ }%
  \textbf{\bibinfo {volume} {331}},\ \bibinfo {pages} {1689--1692} (\bibinfo
  {year} {1994})\BibitemShut{NoStop}%
\bibitem{DAH12b}%
  \BibitemOpen
  \bibfield{author}{%
  \bibinfo {author} {\bibfnamefont{M.~A.}\ \bibnamefont{Dahlem}}\ and\ \bibinfo
  {author} {\bibfnamefont{T.~M.}\ \bibnamefont{Isele}},\ }%
  \bibfield{title}{%
  \enquote{\bibinfo {title} {Transient localized wave patterns and their
  application to migraine},}\ }%
  \bibfield{journal}{%
  \bibinfo {journal} {J. Math. Neurosci}}%
   (\bibinfo {year} {2011})\BibitemShut{NoStop}%
\bibitem{DAH04b}%
  \BibitemOpen
  \bibfield{author}{%
  \bibinfo {author} {\bibfnamefont{M.~A.}\ \bibnamefont{Dahlem}}\ and\ \bibinfo
  {author} {\bibfnamefont{S.~C.}\ \bibnamefont{M{\"u}ller}},\ }%
  \bibfield{title}{%
  \enquote{\bibinfo {title} {{Reaction-diffusion waves in neuronal tissue and
  the window of cortical excitability}},}\ }%
  \bibfield{journal}{%
  \bibinfo {journal} {Ann. Phys.}\ }%
  \textbf{\bibinfo {volume} {13}},\ \bibinfo {pages} {442--449} (\bibinfo
  {year} {2004})\BibitemShut{NoStop}%
\bibitem{DAH08d}%
  \BibitemOpen
  \bibfield{author}{%
  \bibinfo {author} {\bibfnamefont{M.~A.}\ \bibnamefont{Dahlem}}\ and\ \bibinfo
  {author} {\bibfnamefont{N.}~\bibnamefont{Hadjikhani}},\ }%
  \bibfield{title}{%
  \enquote{\bibinfo {title} {Migraine aura: retracting particle-like waves in
  weakly susceptible cortex},}\ }%
  \bibfield{journal}{%
  \Doi{doi:10.1371/journal.pone.0005007}{\bibinfo {journal} {PLOS ONE}}\ }%
  \textbf{\bibinfo {volume} {4}},\ \bibinfo {pages} {e5007} (\bibinfo {year}
  {2009})\BibitemShut{NoStop}%
\bibitem{JAM99}%
  \BibitemOpen
  \bibfield{author}{%
  \bibinfo {author} {\bibfnamefont{M.~F.}\ \bibnamefont{James}}, \bibinfo
  {author} {\bibfnamefont{M.~I.}\ \bibnamefont{Smith}}, \bibinfo {author}
  {\bibfnamefont{K.H.}\ \bibnamefont{Bockhorst}}, \bibinfo {author}
  {\bibfnamefont{L.D.}\ \bibnamefont{Hall}}, \bibinfo {author}
  {\bibfnamefont{G.C.}\ \bibnamefont{Houston}}, \bibinfo {author}
  {\bibfnamefont{N.G.}\ \bibnamefont{Papadakis}}, \bibinfo {author}
  {\bibfnamefont{J.M.}\ \bibnamefont{Smith}}, \bibinfo {author}
  {\bibfnamefont{A.J.}\ \bibnamefont{Williams}}, \bibinfo {author}
  {\bibfnamefont{D.}~\bibnamefont{Xing}}, \bibinfo {author}
  {\bibfnamefont{A.A.}\ \bibnamefont{Parsons}}, \bibinfo {author}
  {\bibfnamefont{C.L.}\ \bibnamefont{Huang}},\ and\ \bibinfo {author}
  {\bibfnamefont{T.A.}\ \bibnamefont{Carpenter}},\ }%
  \bibfield{title}{%
  \enquote{\bibinfo {title} {{{C}ortical spreading depression in the
  gyrencephalic feline brain studied by magnetic resonance imaging}},}\ }%
  \bibfield{journal}{%
  \bibinfo {journal} {J. Physiol. (Lond.)}\ }%
  \textbf{\bibinfo {volume} {519 Pt 2}},\ \bibinfo {pages} {415--425} (\bibinfo
  {year} {1999})\BibitemShut{NoStop}%
\bibitem{AKH08}%
  \BibitemOpen
  \bibfield{author}{%
  \bibinfo {author} {\bibfnamefont{Nail}\ \bibnamefont{Akhmediev}}\ and\
  \bibinfo {author} {\bibfnamefont{Adrian}\ \bibnamefont{Ankiewicz}},\ }%
  \emph{\bibinfo {title} {Dissipative solitons: from optics to biology and
  medicine}},\ Vol.\ \bibinfo {volume} {751}\ (\bibinfo {publisher}
  {Springer},\ \bibinfo {year} {2008})\BibitemShut{NoStop}%
\bibitem{VIN07}%
  \BibitemOpen
  \bibfield{author}{%
  \bibinfo {author} {\bibfnamefont{M.}~\bibnamefont{Vincent}}\ and\ \bibinfo
  {author} {\bibfnamefont{N.}~\bibnamefont{Hadjikhani}},\ }%
  \bibfield{title}{%
  \enquote{\bibinfo {title} {{{M}igraine aura and related phenomena: beyond
  scotomata and scintillations}},}\ }%
  \bibfield{journal}{%
  \bibinfo {journal} {Cephalalgia}\ }%
  \textbf{\bibinfo {volume} {27}},\ \bibinfo {pages} {1368--1377} (\bibinfo
  {year} {2007})\BibitemShut{NoStop}%
\bibitem{DAH12a}%
  \BibitemOpen
  \bibfield{author}{%
  \bibinfo {author} {\bibfnamefont{M.A.}\ \bibnamefont{Dahlem}}\ and\ \bibinfo
  {author} {\bibfnamefont{J.}~\bibnamefont{Tusch}},\ }%
  \bibfield{title}{%
  \enquote{\bibinfo {title} {Predicted selective increase of cortical
  magnification due to cortical folding},}\ }%
  \bibfield{journal}{%
  \bibinfo {journal} {J.~Math. Neurosci.}\ }%
  \textbf{\bibinfo {volume} {2}},\ \bibinfo {pages} {14} (\bibinfo {year}
  {2012})\BibitemShut{NoStop}%
\bibitem{DAH00a}%
  \BibitemOpen
  \bibfield{author}{%
  \bibinfo {author} {\bibfnamefont{M.~A.}\ \bibnamefont{Dahlem}}, \bibinfo
  {author} {\bibfnamefont{R.}~\bibnamefont{Engelmann}}, \bibinfo {author}
  {\bibfnamefont{S.}~\bibnamefont{L{\"o}wel}},\ and\ \bibinfo {author}
  {\bibfnamefont{S.~C.}\ \bibnamefont{M{\"u}ller}},\ }%
  \bibfield{title}{%
  \enquote{\bibinfo {title} {{Does the migraine aura reflect cortical
  organization}},}\ }%
  \bibfield{journal}{%
  \bibinfo {journal} {Eur. J. Neurosci.}\ }%
  \textbf{\bibinfo {volume} {12}},\ \bibinfo {pages} {767--770} (\bibinfo
  {year} {2000})\BibitemShut{NoStop}%
\bibitem{GRA63}%
  \BibitemOpen
  \bibfield{author}{%
  \bibinfo {author} {\bibfnamefont{B.}~\bibnamefont{Grafstein}},\ }%
  \enquote{\bibinfo {title} {Neural release of potassium during spreading
  depression.}.}\ in\ \emph{\bibinfo {booktitle} {Brain Function. Cortical
  Excitability and Steady Potentials}},\ \bibinfo {editor} {edited by\ \bibinfo
  {editor} {\bibfnamefont{M.~A.~B.}\ \bibnamefont{Brazier}}}\ (\bibinfo
  {publisher} {University of California Press},\ \bibinfo {address}
  {Berkeley},\ \bibinfo {year} {1963})\ pp.\ \bibinfo {pages}
  {87--124}\BibitemShut{NoStop}%
\bibitem{HEI77}%
  \BibitemOpen
  \bibfield{author}{%
  \bibinfo {author} {\bibfnamefont{U.}~\bibnamefont{Heinemann}}\ and\ \bibinfo
  {author} {\bibfnamefont{H.~D.}\ \bibnamefont{Lux}},\ }%
  \bibfield{title}{%
  \enquote{\bibinfo {title} {{{C}eiling of stimulus induced rises in
  extracellular potassium concentration in the cerebral cortex of cat}},}\ }%
  \bibfield{journal}{%
  \bibinfo {journal} {Brain Res.}\ }%
  \textbf{\bibinfo {volume} {120}},\ \bibinfo {pages} {231--249} (\bibinfo
  {year} {1977})\BibitemShut{NoStop}%
\bibitem{DAH08}%
  \BibitemOpen
  \bibfield{author}{%
  \bibinfo {author} {\bibfnamefont{M.~A.}\ \bibnamefont{Dahlem}}, \bibinfo
  {author} {\bibfnamefont{F.~M.}\ \bibnamefont{Schneider}},\ and\ \bibinfo
  {author} {\bibfnamefont{E.}~\bibnamefont{Sch{\"o}ll}},\ }%
  \bibfield{title}{%
  \enquote{\bibinfo {title} {Failure of feedback as a putative common mechanism
  of spreading depolarizations in migraine and stroke},}\ }%
  \bibfield{journal}{%
  \Doi{10.1063/1.2937120}{\bibinfo {journal} {Chaos}}\ }%
  \textbf{\bibinfo {volume} {18}},\ \bibinfo {pages} {026110} (\bibinfo {year}
  {2008})\BibitemShut{NoStop}%
\bibitem{BEL95}%
  \BibitemOpen
  \bibfield{author}{%
  \bibinfo {author} {\bibfnamefont{J.}~\bibnamefont{Belair}}, \bibinfo {author}
  {\bibfnamefont{L.}~\bibnamefont{Glass}}, \bibinfo {author}
  {\bibfnamefont{U.}~\bibnamefont{An~Der~Heiden}},\ and\ \bibinfo {author}
  {\bibfnamefont{J.}~\bibnamefont{Milton}},\ }%
  \bibfield{title}{%
  \enquote{\bibinfo {title} {{D}ynamical disease: {I}dentification, temporal
  aspects and treatment strategies of human illness},}\ }%
  \bibfield{journal}{%
  \bibinfo {journal} {Chaos}\ }%
  \textbf{\bibinfo {volume} {5}},\ \bibinfo {pages} {1--7} (\bibinfo {year}
  {1995})\BibitemShut{NoStop}%
\bibitem{MAG12}%
  \BibitemOpen
  \bibfield{author}{%
  \bibinfo {author} {\bibfnamefont{D.}~\bibnamefont{Magis}}\ and\ \bibinfo
  {author} {\bibfnamefont{J.}~\bibnamefont{Schoenen}},\ }%
  \bibfield{title}{%
  \enquote{\bibinfo {title} {{{A}dvances and challenges in neurostimulation for
  headaches}},}\ }%
  \bibfield{journal}{%
  \bibinfo {journal} {Lancet Neurol}\ }%
  \textbf{\bibinfo {volume} {11}},\ \bibinfo {pages} {708--719} (\bibinfo
  {year} {2012})\BibitemShut{NoStop}%
\bibitem{VEL03}%
  \BibitemOpen
  \bibfield{author}{%
  \bibinfo {author} {\bibfnamefont{F.}\ \bibnamefont{Veloso}}, \bibinfo
  {author} {\bibfnamefont{K.}\ \bibnamefont{Kumar}},\ and\ \bibinfo
  {author} {\bibfnamefont{C.}\ \bibnamefont{Toth}},\ }%
  \bibfield{title}{%
  \enquote{\bibinfo {title} {Headache secondary to deep brain implantation},}\
  }%
  \bibfield{journal}{%
  \bibinfo {journal} {Headache: The Journal of Head and Face Pain}\ }%
  \textbf{\bibinfo {volume} {38}},\ \bibinfo {pages} {507--515} (\bibinfo
  {year} {2003})\BibitemShut{NoStop}%
\bibitem{RAS05}%
  \BibitemOpen
  \bibfield{author}{%
  \bibinfo {author} {\bibfnamefont{Neil~H}\ \bibnamefont{Raskin}}, \bibinfo
  {author} {\bibfnamefont{Yoshio}\ \bibnamefont{Hosobuchi}},\ and\ \bibinfo
  {author} {\bibfnamefont{Sharon}\ \bibnamefont{Lamb}},\ }%
  \bibfield{title}{%
  \enquote{\bibinfo {title} {Headache may arise from perturbation of brain},}\
  }%
  \bibfield{journal}{%
  \bibinfo {journal} {Headache: The Journal of Head and Face Pain}\ }%
  \textbf{\bibinfo {volume} {27}},\ \bibinfo {pages} {416--420} (\bibinfo
  {year} {2005})\BibitemShut{NoStop}%
\bibitem{SCH10h}%
  \BibitemOpen
  \bibfield{author}{%
  \bibinfo {author} {\bibfnamefont{S.~J.}\ \bibnamefont{Schiff}},\ }%
  \bibfield{title}{%
  \enquote{\bibinfo {title} {{{T}owards model-based control of {P}arkinson's
  disease}},}\ }%
  \bibfield{journal}{%
  \bibinfo {journal} {Phil. Trans. R. Soc. A}\ }%
  \textbf{\bibinfo {volume} {368}},\ \bibinfo {pages} {2269--2308} (\bibinfo
  {year} {2010})\BibitemShut{NoStop}%
\bibitem{RUB12}%
  \BibitemOpen
  \bibfield{author}{%
  \bibinfo {author} {\bibfnamefont{J.~E.}\ \bibnamefont{Rubin}}, \bibinfo
  {author} {\bibfnamefont{C.~C.}\ \bibnamefont{McIntyre}}, \bibinfo {author}
  {\bibfnamefont{R.~S.}\ \bibnamefont{Turner}},\ and\ \bibinfo {author}
  {\bibfnamefont{T.}~\bibnamefont{Wichmann}},\ }%
  \bibfield{title}{%
  \enquote{\bibinfo {title} {Basal ganglia activity patterns in parkinsonism
  and computational modeling of their downstream effects},}\ }%
  \bibfield{journal}{%
  \bibinfo {journal} {Eur. J.~Neurosci.}\ }%
  \textbf{\bibinfo {volume} {36}},\ \bibinfo {pages} {2213--2228} (\bibinfo
  {year} {2012})\BibitemShut{NoStop}%
\bibitem{PAE12}%
  \BibitemOpen
  \bibfield{author}{%
  \bibinfo {author} {\bibfnamefont{K.}~\bibnamefont{Paemeleire}}\ and\ \bibinfo
  {author} {\bibfnamefont{A.~M.}\ \bibnamefont{Goodman}},\ }%
  \bibfield{title}{%
  \enquote{\bibinfo {title} {{{R}esults of a patient survey for an implantable
  neurostimulator to treat migraine headaches}},}\ }%
  \bibfield{journal}{%
  \bibinfo {journal} {J.~Headache Pain}\ }%
  \textbf{\bibinfo {volume} {13}},\ \bibinfo {pages} {239--241} (\bibinfo
  {year} {2012})\BibitemShut{NoStop}%
\bibitem{TEP09}%
  \BibitemOpen
  \bibfield{author}{%
  \bibinfo {author} {\bibfnamefont{Stewart~J.}\ \bibnamefont{Tepper}}, \bibinfo
  {author} {\bibfnamefont{Ali}\ \bibnamefont{Rezai}}, \bibinfo {author}
  {\bibfnamefont{Samer}\ \bibnamefont{Narouze}}, \bibinfo {author}
  {\bibfnamefont{Charles}\ \bibnamefont{Steiner}}, \bibinfo {author}
  {\bibfnamefont{Pouya}\ \bibnamefont{Mohajer}},\ and\ \bibinfo {author}
  {\bibfnamefont{Mehdi}\ \bibnamefont{Ansarinia}},\ }%
  \bibfield{title}{%
  \enquote{\bibinfo {title} {Acute treatment of intractable migraine with
  sphenopalatine ganglion electrical stimulation},}\ }%
  \bibfield{journal}{%
  \Doi{10.1111/j.1526-4610.2009.01451.x}{\bibinfo {journal} {Headache}}\ }%
  \textbf{\bibinfo {volume} {49}},\ \bibinfo {pages} {983--989} (\bibinfo
  {year} {2009}),\ ISSN \bibinfo {issn} {1526-4610}\BibitemShut{NoStop}%
\bibitem{POP03}%
  \BibitemOpen
  \bibfield{author}{%
  \bibinfo {author} {\bibfnamefont{C.A.}\ \bibnamefont{Popeney}}\ and\
  \bibinfo {author} {\bibfnamefont{K.M.}\ \bibnamefont{Alo}},\ }%
  \bibfield{title}{%
  \enquote{\bibinfo {title} {{{P}eripheral neurostimulation for the treatment
  of chronic, disabling transformed migraine}},}\ }%
  \bibfield{journal}{%
  \bibinfo {journal} {Headache}\ }%
  \textbf{\bibinfo {volume} {43}},\ \bibinfo {pages} {369--375} (\bibinfo
  {year} {2003})\BibitemShut{NoStop}%
\bibitem{SCH07i}%
  \BibitemOpen
  \bibfield{author}{%
  \bibinfo {author} {\bibfnamefont{T.J.}~\bibnamefont{Schwedt}}, \bibinfo
  {author} {\bibfnamefont{D.W.}~\bibnamefont{Dodick}}, \bibinfo {author}
  {\bibfnamefont{J.}~\bibnamefont{Hentz}}, \bibinfo {author}
  {\bibfnamefont{T.L.}~\bibnamefont{Trentman}},\ and\ \bibinfo {author}
  {\bibfnamefont{R.S.}~\bibnamefont{Zimmerman}},\ }%
  \bibfield{title}{%
  \enquote{\bibinfo {title} {Occipital nerve stimulation for chronic
  headache---long-term safety and efficacy},}\ }%
  \bibfield{journal}{%
  \bibinfo {journal} {Cephalalgia}\ }%
  \textbf{\bibinfo {volume} {27}},\ \bibinfo {pages} {153--157} (\bibinfo
  {year} {2007})\BibitemShut{NoStop}%
\bibitem{MAG07}%
  \BibitemOpen
  \bibfield{author}{%
  \bibinfo {author} {\bibfnamefont{D.}~\bibnamefont{Magis}}, \bibinfo {author}
  {\bibfnamefont{M.}~\bibnamefont{Allena}}, \bibinfo {author}
  {\bibfnamefont{M.}~\bibnamefont{Bolla}}, \bibinfo {author}
  {\bibfnamefont{V.}~\bibnamefont{De~Pasqua}}, \bibinfo {author}
  {\bibfnamefont{J.~M.}\ \bibnamefont{Remacle}},\ and\ \bibinfo {author}
  {\bibfnamefont{J.}~\bibnamefont{Schoenen}},\ }%
  \bibfield{title}{%
  \enquote{\bibinfo {title} {{{O}ccipital nerve stimulation for drug-resistant
  chronic cluster headache: a prospective pilot study}},}\ }%
  \bibfield{journal}{%
  \bibinfo {journal} {Lancet Neurol}\ }%
  \textbf{\bibinfo {volume} {6}},\ \bibinfo {pages} {314--321} (\bibinfo {year}
  {2007})\BibitemShut{NoStop}%
\bibitem{MAG11}%
  \BibitemOpen
  \bibfield{author}{%
  \bibinfo {author} {\bibfnamefont{D.}~\bibnamefont{Magis}}, \bibinfo {author}
  {\bibfnamefont{P.Y.}\ \bibnamefont{Gerardy}}, \bibinfo {author}
  {\bibfnamefont{J.M.}\ \bibnamefont{Remacle}},\ and\ \bibinfo {author}
  {\bibfnamefont{J.}~\bibnamefont{Schoenen}},\ }%
  \bibfield{title}{%
  \enquote{\bibinfo {title} {{{S}ustained effectiveness of occipital nerve
  stimulation in drug-resistant chronic cluster headache}},}\ }%
  \bibfield{journal}{%
  \bibinfo {journal} {Headache}\ }%
  \textbf{\bibinfo {volume} {51}},\ \bibinfo {pages} {1191--1201} (\bibinfo
  {year} {2011})\BibitemShut{NoStop}%
\bibitem{SCH13d}%
  \BibitemOpen
  \bibfield{author}{%
  \bibinfo {author} {\bibfnamefont{J.}\ \bibnamefont{Schoenen}}, \bibinfo
  {author} {\bibfnamefont{B.}\ \bibnamefont{Vandersmissen}}, \bibinfo
  {author} {\bibfnamefont{S.}\ \bibnamefont{Jeangette}}, \bibinfo
  {author} {\bibfnamefont{L.}\ \bibnamefont{Herroelen}}, \bibinfo {author}
  {\bibfnamefont{M.}\ \bibnamefont{Vandenheede}}, \bibinfo {author}
  {\bibfnamefont{P.}\ \bibnamefont{G{\'e}rard}},\ and\ \bibinfo {author}
  {\bibfnamefont{D.}\ \bibnamefont{Magis}},\ }%
  \bibfield{title}{%
  \enquote{\bibinfo {title} {Migraine prevention with a supraorbital
  transcutaneous stimulator. {A} randomized controlled trial},}\ }%
  \bibfield{journal}{%
  \bibinfo {journal} {Neurology}\ }%
  \textbf{\bibinfo {volume} {80}},\ \bibinfo {pages} {697--704} (\bibinfo
  {year} {2013})\BibitemShut{NoStop}%
\bibitem{LIP10}%
  \BibitemOpen
  \bibfield{author}{%
  \bibinfo {author} {\bibfnamefont{R.B.}\ \bibnamefont{Lipton}}, \bibinfo
  {author} {\bibfnamefont{D.W.}\ \bibnamefont{Dodick}}, \bibinfo {author}
  {\bibfnamefont{S.D.}\ \bibnamefont{Silberstein}}, \bibinfo {author}
  {\bibfnamefont{J.R.}\ \bibnamefont{Saper}}, \bibinfo {author}
  {\bibfnamefont{S.K.}\ \bibnamefont{Aurora}}, \bibinfo {author}
  {\bibfnamefont{S.H.}\ \bibnamefont{Pearlman}}, \bibinfo {author}
  {\bibfnamefont{R.E.}\ \bibnamefont{Fischell}}, \bibinfo {author}
  {\bibfnamefont{P.L.}\ \bibnamefont{Ruppel}},\ and\ \bibinfo {author}
  {\bibfnamefont{P.J.}\ \bibnamefont{Goadsby}},\ }%
  \bibfield{title}{%
  \enquote{\bibinfo {title} {{S}ingle-pulse transcranial magnetic stimulation
  for acute treatment of migraine with aura: a randomised, double-blind,
  parallel-group, sham-controlled trial},}\ }%
  \bibfield{journal}{%
  \bibinfo {journal} {Lancet. Neurol.}\ }%
  \textbf{\bibinfo {volume} {9}},\ \bibinfo {pages} {373--380} (\bibinfo {year}
  {2010})\BibitemShut{NoStop}%
\bibitem{DAS12}%
  \BibitemOpen
  \bibfield{author}{%
  \bibinfo {author} {\bibfnamefont{A.~F.}\ \bibnamefont{Dasilva}}, \bibinfo
  {author} {\bibfnamefont{M.~E.}\ \bibnamefont{Mendonca}}, \bibinfo {author}
  {\bibfnamefont{S.}~\bibnamefont{Zaghi}}, \bibinfo {author}
  {\bibfnamefont{M.}~\bibnamefont{Lopes}}, \bibinfo {author}
  {\bibfnamefont{M.~F.}\ \bibnamefont{Dossantos}}, \bibinfo {author}
  {\bibfnamefont{E.~L.}\ \bibnamefont{Spierings}}, \bibinfo {author}
  {\bibfnamefont{Z.}~\bibnamefont{Bajwa}}, \bibinfo {author}
  {\bibfnamefont{A.}~\bibnamefont{Datta}}, \bibinfo {author}
  {\bibfnamefont{M.}~\bibnamefont{Bikson}},\ and\ \bibinfo {author}
  {\bibfnamefont{F.}~\bibnamefont{Fregni}},\ }%
  \bibfield{title}{%
  \enquote{\bibinfo {title} {{t{D}{C}{S}-induced analgesia and electrical
  fields in pain-related neural networks in chronic migraine}},}\ }%
  \bibfield{journal}{%
  \bibinfo {journal} {Headache}\ }%
  \textbf{\bibinfo {volume} {52}},\ \bibinfo {pages} {1283--1295} (\bibinfo
  {year} {2012})\BibitemShut{NoStop}%
\bibitem{ANT11a}%
  \BibitemOpen
  \bibfield{author}{%
  \bibinfo {author} {\bibfnamefont{A.}~\bibnamefont{Antal}}, \bibinfo {author}
  {\bibfnamefont{N.}~\bibnamefont{Kriener}}, \bibinfo {author}
  {\bibfnamefont{N.}~\bibnamefont{Lang}}, \bibinfo {author}
  {\bibfnamefont{K.}~\bibnamefont{Boros}},\ and\ \bibinfo {author}
  {\bibfnamefont{W.}~\bibnamefont{Paulus}},\ }%
  \bibfield{title}{%
  \enquote{\bibinfo {title} {{{C}athodal transcranial direct current
  stimulation of the visual cortex in the prophylactic treatment of
  migraine}},}\ }%
  \bibfield{journal}{%
  \bibinfo {journal} {Cephalalgia}\ }%
  \textbf{\bibinfo {volume} {31}},\ \bibinfo {pages} {820--828} (\bibinfo
  {year} {2011})\BibitemShut{NoStop}%
\bibitem{MIU07}%
  \BibitemOpen
  \bibfield{author}{%
  \bibinfo {author} {\bibfnamefont{R.~M.}\ \bibnamefont{Miura}}, \bibinfo
  {author} {\bibfnamefont{H.}~\bibnamefont{Huang}},\ and\ \bibinfo {author}
  {\bibfnamefont{J.~J.}\ \bibnamefont{Wylie}},\ }%
  \bibfield{title}{%
  \enquote{\bibinfo {title} {Cortical spreading depression: An enigma},}\ }%
  \bibfield{journal}{%
  \bibinfo {journal} {Eur. Phys. J. Spec. Top.}\ }%
  \textbf{\bibinfo {volume} {147}},\ \bibinfo {pages} {287--302} (\bibinfo
  {year} {2007})\BibitemShut{NoStop}%
\bibitem{MAY06a}%
  \BibitemOpen
  \bibfield{author}{%
  \bibinfo {author} {\bibfnamefont{A.}~\bibnamefont{May}},\ }%
  \bibfield{title}{%
  \enquote{\bibinfo {title} {{{A} review of diagnostic and functional imaging
  in headache}},}\ }%
  \bibfield{journal}{%
  \bibinfo {journal} {J Headache Pain}\ }%
  \textbf{\bibinfo {volume} {7}},\ \bibinfo {pages} {174--184} (\bibinfo {year}
  {2006})\BibitemShut{NoStop}%
\bibitem{PIE03}%
  \BibitemOpen
  \bibfield{author}{%
  \bibinfo {author} {\bibfnamefont{D.}~\bibnamefont{Pietrobon}}\ and\ \bibinfo
  {author} {\bibfnamefont{J.}~\bibnamefont{Striessnig}},\ }%
  \bibfield{title}{%
  \enquote{\bibinfo {title} {{{N}eurobiology of migraine}},}\ }%
  \bibfield{journal}{%
  \bibinfo {journal} {Nat. Rev. Neurosci.}\ }%
  \textbf{\bibinfo {volume} {4}},\ \bibinfo {pages} {386--398} (\bibinfo {year}
  {2003})\BibitemShut{NoStop}%
\end{thebibliography}
\end{document}